\documentclass[journal,twoside]{IEEEtran}

\ifdefined\pdfmapfile
  \pdfmapfile{+mdsymbol.map}
  \pdfmapfile{+dstroke.map}
  \pdfmapfile{+boondox.map}
\fi

\usepackage{enumerate}
\usepackage{supertabular,booktabs}

\usepackage{tabularx}
\usepackage{array}

\usepackage{amsfonts}
\usepackage{color}

\usepackage{cite}

\usepackage{mathtools}
\mathtoolsset{showonlyrefs=true}

\usepackage{array}

\usepackage{multirow}

\usepackage{stackrel}

\usepackage{tikz}

\usetikzlibrary{shapes,arrows,decorations.markings,positioning}

\tikzstyle{plant} = [draw, fill=red!5, rounded rectangle, 
    minimum height=3em]
\tikzstyle{block} = [draw, fill=blue!5, rectangle, 
    minimum height=3em, minimum width = 15mm]
\tikzstyle{tap} = [draw, fill=olive!10, rectangle, minimum height=3em]
\tikzstyle{sum} = [draw, fill=yellow!10, circle, node distance=1cm]
\tikzstyle{pinstyle} = [pin edge={to-,thick,black}]

\tikzstyle{BitPipe} = [thick, decoration={markings,mark=at position
   1 with {\arrow[semithick]{triangle 60}}},
   double distance=1.4pt, shorten >= 5.5pt,
   preaction = {decorate},
   postaction = {draw,line width=1.4pt, black, shorten >= 4.5pt}]

\usetikzlibrary{chains,shapes.multipart}
\tikzstyle{FIFO} = [rectangle split, rectangle split parts=3, draw, rectangle split horizontal,minimum height=3em,text height=1.5em,text depth=1em,on chain,inner ysep=0pt]

\usetikzlibrary{fit,backgrounds}

\tikzstyle{coord} = [coordinate]
\tikzstyle{gain} = [draw, fill=red!5, regular polygon, regular polygon sides=3, shape border rotate=-90]

\tikzset{mwe/.style={minimum size=0.5em}}

\usepackage{hyperref}

\usepackage{epsfig, graphicx, psfrag}

\usepackage[mathcal]{euscript}
\usepackage[cal=boondox]{mathalfa}

\newcommand{\khina}{{A.~Khina}}
\newcommand{\MyStudent}[1]{#1}

\usepackage{amsthm}

\theoremstyle{plain}
\newtheorem{thm}{Theorem}
\newtheorem{lem}{Lemma}
\newtheorem{cor}{Corollary}

\theoremstyle{definition}
\newtheorem{defn}{Definition}
\newtheorem*{defn*}{Definition}
\newtheorem*{scheme*}{Scheme}
\newtheorem{example}{Example}

\theoremstyle{remark}
\newtheorem{remark}{Remark}

\numberwithin{thm}{section}
\numberwithin{lem}{section}
\numberwithin{claim}{section}
\numberwithin{assert}{section}
\numberwithin{cor}{section}
\numberwithin{prop}{section}
\numberwithin{defn}{section}
\numberwithin{remark}{section}
\numberwithin{algo}{section}
\numberwithin{scheme}{section}
\numberwithin{example}{section}

\providecommand{\thmref}[1]{Theorem~\ref{#1}}

\providecommand{\exref}[1]{Example~\ref{#1}}
\providecommand{\defref}[1]{Definition~\ref{#1}}

\providecommand{\secref}[1]{Section~\ref{#1}}
\providecommand{\lemref}[1]{Lemma~\ref{#1}}

\providecommand{\remref}[1]{Remark~\ref{#1}}
\providecommand{\figref}[1]{Figure~\ref{#1}}
\providecommand{\corref}[1]{Corollary~\ref{#1}}

\providecommand{\appref}[1]{Appendix~\ref{#1}}

\providecommand{\tabref}[1]{Table~\ref{#1}}

\providecommand{\footref}[1]{\footnotemark[\ref{#1}]}

\usepackage{cancel}

\newcommand{\viz}{viz.}

\newcommand{\reals}{\mathbb{R}}
\newcommand{\ints}{\mathbb{Z}}

\newcommand{\nats}{\mathbb{N}}

\newcommand{\firstnats}[1]{\lrc{1,2,\ldots,#1}}

\newcommand{\floor}[1]{\left\lfloor{#1}\right\rfloor}

\let\limsup\relax
\DeclareMathOperator*{\limsup}{\overline{lim}}
\let\liminf\relax
\DeclareMathOperator*{\liminf}{\underline{lim}}

\newcommand{\X}{Y}
\newcommand{\Y}{X}
\newcommand{\x}{y}
\newcommand{\y}{x}

\newcommand{\e}{\mathrm{e}}

\newcommand{\old}[1]{}
\newcommand{\rem}[1]{}

\newcommand{\eps}{{\epsilon}}

\newcommand{\MMSE}[2]{\mathrm{MMSE}\lrpm{#1}{#2}}
\newcommand{\LMMSE}[2]{\mathrm{LMMSE}\lrpm{#1}{#2}}

\newcommand{\hx}{{\hat{x}}}

\newcommand{\tX}{\tilde{\X}}

\newcommand{\ty}{\tilde \y}

\newcommand{\tY}{\tilde \Y}

\newcommand{\cS}{{\mathcal S}}

\newcommand{\abs}[1]{\left| #1 \right|}

\providecommand{\e}{{\rm e}}

\providecommand{\norm}[1]{\left\| #1 \right\|}
\providecommand{\normE}[1]{\norm{#1}_{L^2}}
\providecommand{\normLone}[1]{\norm{#1}_{L^1}}
\providecommand{\normEr}[1]{\norm{#1}_{L^r}}
\providecommand{\normInf}[1]{\norm{#1}_{\infty}}
\providecommand{\inner}[2]{\ensuremath{\left\langle #1, #2 \right\rangle}}
\providecommand{\innerE}[2]{\left\langle #1, #2 \right\rangle_{L^2}}

\newcommand{\dgamma}{\mathrm{d}\gamma}

\providecommand{\var}[1]{\mathrm{Var}\left( #1 \right)}
\providecommand{\Var}[1]{\ensuremath{\mathrm{Var} \left( #1 \right)}}

\providecommand{\cov}[1]{\ensuremath{\mathrm{Cov} \lrp{#1}}}
\providecommand{\Cov}[2]{\ensuremath{\mathrm{Cov} \lrp{#1, #2}}}

\usepackage{pifont}
\usepackage{mdsymbol}
\newcommand{\indep}{\upvDash}

\usepackage{dsfont}

\providecommand{\qqquad}{\quad \qquad}

\providecommand{\trace}{\mathrm{trace}}
\providecommand{\det}{\mathrm{det}}
\providecommand{\adj}{\mathrm{adj}}
\providecommand{\half}{\frac{1}{2}}

\providecommand{\trace}{\text{trace}}

\providecommand{\half}{\frac{1}{2}}

\makeatletter
\newcommand{\bigger}{\bBigg@{3.2}}
\newcommand{\vast}{\bBigg@{4}}
\newcommand{\Vast}{\bBigg@{5}}
\makeatother

\newcommand{\defeq}{\triangleq}

\DeclareMathOperator{\opI}{I}
\DeclareMathOperator{\oph}{h}
\DeclareMathOperator{\opH}{H}

\providecommand{\KL}[2]{{\mathrm{D}_\mathrm{KL} \left( #1 \middle\| #2 \right)}}
\providecommand{\JSD}[2]{{\mathrm{D}_\mathrm{JS} \left( #1 \middle\| #2 \right)}}
\providecommand{\MI}[2]{{\opI \left( #1 ; #2 \right)}}

\renewcommand{\H}[1]{{\opH \left( #1 \right)}}
\providecommand{\h}[1]{{\oph \left( #1 \right)}}

\providecommand{\E}[1]{\mathrm{E} \left[ #1 \right]}
\providecommand{\Esqr}[1]{\mathrm{E}^2 \left[ #1 \right]}
\providecommand{\CE}[2]{\mathrm{E} \left[ #1 \middle| #2 \right]}

\newcommand{\PR}[1]{\mathrm{P}\left( #1 \right)}

\providecommand{\hg}{\hat{g}}

\providecommand{\hx}{\hat{x}}

\providecommand{\hY}{\hat{\Y}}

\providecommand{\cE}{\mathcal{E}}
\providecommand{\cF}{\mathcal{F}}

\providecommand{\cS}{\mathcal{S}}

\providecommand{\cX}{\mathcal{\X}}
\providecommand{\cY}{\mathcal{\Y}}

\newcommand{\xto}{\xrightarrow}
\newcommand{\stochto}{\xto[n \to \infty]}

\newcommand{\indicator}[1]{\mathds{1}\left\{ #1 \right\}}

\providecommand{\lrp}[1]{\left( #1 \right)}

\providecommand{\lrs}[1]{\left[ #1 \right]}
\providecommand{\lrc}[1]{\left\{ #1 \right\}}
\providecommand{\lrpm}[2]{\left( #1 \middle| #2 \right)}

\providecommand{\lrcm}[2]{\left\{ #1 \middle| #2 \right\}}

\newcommand{\markov}{\mathrel{-\mkern-6mu\circ\mkern-6mu-}}
\providecommand{\dmarkov}{\stackrel{\mathrm{d}}{\markov}}

\DeclareMathOperator{\sign}{sign}
\providecommand{\Lim}{\lim\limits}
\newcommand{\seq}[1]{\lrc{#1}_{n=1}^\infty}
\newcommand{\cost}[2]{c \lrp{#1, #2}}
\newcommand{\ocost}[2]{\bar{c}_\mathrm{opt} \lrpm{#1}{#2}}
\newcommand{\oy}{\bar{\y}}
\newcommand{\limn}{\lim_{n \to \infty}}
\newcommand{\Limn}{\Lim_{n \to \infty}}

\hypersetup{hidelinks,pdftitle={Markovian Continuity of the MMSE},pdfauthor={Elad Domanovitz and Anatoly Khina}}

\begin{document}
\allowdisplaybreaks

\title{Markovian Continuity of the MMSE}

\author{Elad Domanovitz,~\IEEEmembership{Member,~IEEE}, and Anatoly Khina,~\IEEEmembership{Senior Member,~IEEE}

\thanks{
This work was supported in part by the \textsc{Israel Science Foundation} (grants No.\ 2077/20 and 2216/25) and in part by a grant from the Tel Aviv University Center for AI and Data Science (TAD).}
\thanks{The authors are with the School of Electrical and Computer Engineering, Tel Aviv University, Tel Aviv~6997801, Israel (e-mails: \text{domanovi@eng.tau.ac.il, anatolyk@tau.ac.il}).}
\thanks{\copyright\ 2026 IEEE. Personal use of this material is permitted.
Permission from IEEE must be obtained for all other uses, in any current or
future media, including reprinting/republishing this material for advertising
or promotional purposes, creating new collective works, for resale or
redistribution to servers or lists, or reuse of any copyrighted component of
this work in other works.}
} 

\maketitle

\begin{abstract}
Minimum mean square error (MMSE) estimation is widely used in signal processing, information theory, and related fields. Despite its practical robustness, the MMSE can be discontinuous under standard notions of stochastic convergence. To bridge this gap, we review classical counterexamples to the continuity of the MMSE and identify two obstructions: failure of second-moment convergence and approximating observations carrying more information about the limit estimand than the limit observation does. Motivated by practical acquisition mechanisms, we study MMSE continuity under two natural constraints: (1) continuity of the second moment, and (2) a degradedness (Markov) restriction ensuring that each approximating observation is no more informative than the limit observation is about the limit estimand. Under these conditions, we establish continuity of the MMSE and of the MMSE estimator. We provide complementary semicontinuity results and continuity guarantees in related settings and establish continuity under linear estimation. We further extend the analysis to the families of Bregman divergences and continuous metric and generalized metric cost functions, including the Kullback--Leibler and Jensen--Shannon divergences as special cases.
\end{abstract}


\section{Introduction}
\label{s:intro}


\IEEEPARstart{M}{inimum} mean square error (MMSE) estimation is a cornerstone principle at the core of many fields, including signal processing, information theory, communications, control theory, and machine learning. By seeking the estimator that minimizes the expected squared error between an unknown quantity and its estimate, MMSE provides a clean and powerful optimality criterion that is both mathematically tractable and practically meaningful as the minimizer of the mean power of the estimation error~\cite{VanTrees:Book:2004:PartI:detection-estimation-modulation,Kay:Book:1993:SSP:estimation,Lehmann-Casella:estimation:book2006}. In signal processing and control, MMSE ideas appear in optimal filters such as the Kalman filter, enabling robust estimation and tracking in noisy, dynamic environments~\cite{kalman1960new,Wiener49book,Kolmogorov:stationary_Hilbert,Babook:Estimation,BertsekasControlVol1,Babook:Indefinite}. In machine learning, MMSE corresponds to training models under squared-loss objectives, linking classical estimation theory with modern regression, representation learning, online learning, reinforcement learning, and probabilistic inference \cite{Birkes-Dodge:Regression:Book2011,R-square-regression:PeerJ2021,SuttonBarto:RL:Book,Cesa-Bianchi--Lugosi:online-learning:book2006,Bishop:pattern-recognition:book2006,Murphy:Machine-learning-prob:book2012}. These deep and recurring connections make MMSE a central tool for analyzing, designing, and understanding a broad class of inference and decision systems.

In the context of information theory, MMSE is intimately related to entropy, differential entropy, mutual information, channel capacity and rate--distortion function (RDF).
For a discrete random variable (RV) $X$ with a finite variance, its entropy satisfies~\cite{Guo-Shamai-Verdu:I-MMSE:TIT2005}
\begin{align}
    \H{X} = \half \int_{0}^\infty \MMSE{X}{\sqrt{\gamma} X + N} \dgamma ,
\end{align}
where $\MMSE{\Y}{\X}$ denotes the MMSE in estimating the random parameter $\Y$ from the measurement~$\X$, 
and $N$ is a standard Gaussian RV independent of $X$.

If, instead, $X$ has a probability density function (PDF) and a finite second moment, then its differential entropy can be represented as~\cite{Guo-Shamai-Verdu:I-MMSE:TIT2005,Verdu-Guo:I-MMSE-->EPI:TIT2006}
\begin{align}
    \begin{aligned}
    \h{X} = {}&\half \log(2\pi\e)\\
    &- \half \int_0^\infty \lrs{\frac{1}{1+\gamma} - \MMSE{X}{\sqrt{\gamma} X + N}} \dgamma .
    \end{aligned}
\end{align}
This representation can then be used to derive Lieb's inequality and Shannon's entropy-power inequality~\cite{Verdu-Guo:I-MMSE-->EPI:TIT2006,Rioul:EPI-proofs:TIT2011} as well as a generalization of the latter for a linear transformation of a random vector (RVec) with independent components~\cite{Zamir-Feder:generalized-EPI:TIT1993}. 

Both of these relations are consequences of the celebrated I--MMSE formula of Guo, Shamai and Verd\'u~\cite{Guo-Shamai-Verdu:I-MMSE:TIT2005}, which relates the MMSE and the
mutual information between the input and the output of an additive white Gaussian noise (AWGN) channel:
\begin{align}
    \frac{\mathrm{d}}{\dgamma} \MI{X}{\sqrt{\gamma} X + N} = \half \MMSE{X}{\sqrt{\gamma}X + N}.
\end{align}
This I-MMSE relation was subsequently used to derive: simple proofs for various Gaussian problems, such as the Gaussian multiple-input multiple-output (MIMO) wiretap channel~\cite{Wiretap_BustinEURASIP,Guo-Shamai-Verdu:I-MMSE:NOW2013,Guo-Wu-Shamai-Verdu:MMSE-properties:TIT2011}, and the Gaussian broadcast channel~\cite{Guo-Wu-Shamai-Verdu:MMSE-properties:TIT2011};
a water-filling-like solution for discrete constellations~\cite{Lozano-Tulino-Verdu:waterfilling-constellation:TIT2006}; and an upper bound on the RDF under quadratic loss~\cite{Dorpinghaus-Rungeler-Mathar:RDF-upper-bound:quadratic-loss:IZS2012}.

A parametric representation of the RDF under quadratic loss via the MMSE, unrelated to the I-MMSE formula, was proposed by Merhav~\cite{Merhav:RDF-via-MMSE:TIT2011,Merhav:statistical-physics-and-information-theory:NOW2010}. He used this representation to derive upper and lower bounds on the RDF under quadratic loss and characterize its asymptotic behavior in the high and low distortion limits.

Communication schemes that rely on the MMSE include equalizers and predictors~\cite{Proakis83,CDFE-PartI,CDFE-PartII,ZamirKochmanErez:DPCM:IT,ZamirBook,Barry-Lee-Messerschmitt:Digital-Comm:Book2004,Domanovitz-Khina-Philosof-Kochman:Gaussian:IV:JSAIT2024}, and the analysis and construction of message passing algorithms~\cite{Feng-Venkataramanan-Rush-Samworth:Approximate-message-passing:NOW2022}, \cite[Section~4.12]{Richardson-Urbanke:modern-coding-theory:Book2008}, \cite{Bhattad-Narayanan:MSE-EXIT-chart:TIT2007}.

Such MMSE-based applications and techniques rely primarily on models that require statistical knowledge.
As the required statistics are acquired from finite samples and finite-precision/noisy measurements, continuity of the MMSE and the corresponding estimators is implicitly assumed. 

Unfortunately, despite being considered robust in practice, the MMSE is known to be discontinuous in general. To wit, consider a sequence of pairs of RVs $\seq{\lrp{X_n, Y_n}}$ converging in some standard stochastic sense (in distribution, in probability, in mean square, almost surely) to a pair of RVs $\lrp{X, Y}$. 
Here, $\lrp{X_n, Y_n}$ can represent, e.g., an empirical distribution resulting from a finite sample of size $n$ drawn from the distribution of $\lrp{X, Y}$, a finite-precision variant of $\lrp{X,Y}$ with the machine precision increasing with $n$, or noisy variants of $\lrp{X,Y}$ with diminishing noise power with $n$.
Then, $\seq{\MMSE{\Y_n}{\X_n}}$ does not converge in general to $\MMSE{\Y}{\X}$:
\begin{align}
    \MMSE{\Y_n}{\X_n} \nlongrightarrow \MMSE{\Y}{\X},
\end{align}
where $\MMSE{\Y}{\X}$ denotes the MMSE in estimating the random parameter $\Y$ from the measurement~$\X$. As is common, we will say that the MMSE is continuous/discontinuous in a particular stochastic sense depending on the stochastic convergence sense of the sequence $\seq{\lrp{X_n,Y_n}}$ to $\lrp{X,Y}$. 

Wu and Verd\'u~\cite{Wu-Verdu:MMSE:IT2012}, and Y\"uksel and Linder~\cite{Yuksel-Linder:continuity-control:SICON2012} (see also \cite{Mori-Szekely:4axioms-dependence:Metrika2019}, \cite[Chapter~8.3]{Yuksel-Basar:games:book:2024}) provided concrete counterexamples to the continuity of the MMSE, 
demonstrating that the discontinuity may even be unbounded. They and Hogeboom-Burr~\cite{Hogeboom_thesis,Hogeboom-Yuksel:continuity:SICON2023} further established some sufficient conditions for the semicontinuity and continuity of the MMSE. 
Guarantees for the more abstract setting of convergence of the MMSE with respect to the convergence of the $\sigma$-subalgebras were derived in a series of works~\cite{Boylan:martinagle-equiconvergance:AMS1971,Fetter:conditional-mean-continuity:JMAA1977,Alonso:conditional-mean-continuity-counterexample:JMAA1988,Alonso--Brambila-Paz:Lp-conditional-mean-continuity:JMAA1998}.

However, the continuity of some important scenarios is not guaranteed by the hitherto existing results.
For example:
\begin{itemize}\addtolength\itemsep{.5\baselineskip}
\item 
    \textit{Additive noises.} Let $X,Y,M,N$ be RVs where 
    $\Y$ and $N$ have finite second moments and $M$ is independent of $(X,Y)$.
    Does
    \begin{align}
    \label{eq:additiveNoise}
        \begin{aligned}
        &\lim_{\lrp{\lambda, \gamma} \to \lrp{0,0}} \MMSE{\Y + \gamma N}{\X + \lambda M}
        = \MMSE{\Y}{\X}
        \end{aligned}
    \end{align}
    hold?
\item 
    \textit{Machine precision.} Let $X$ and $Y$ be two RVs such that $\Y$ has finite second moment.
    Denote by $\floor{\cdot}$ the floor operation. 
    Does
        \begin{align}
        \label{eq:machinePrecision}
            \begin{aligned}
            &\lim_{\substack{\lrp{\lambda, \gamma} \to \lrp{0,0} \\ \lambda > 0, \gamma > 0}}
            \MMSE{\floor{\frac{\Y}{\gamma}} \cdot \gamma\,}{\floor{\frac{\X}{\lambda}} \cdot \lambda}
            = \MMSE{\Y}{\X}
            \end{aligned}
        \end{align} 
    hold?
\end{itemize}

In this work, we establish new continuity results for the MMSE, which subsume the aforementioned results. To that end, we first review existing counterexamples to and guarantees of the continuity of the MMSE in \secref{s:existing}.
We identify common traits in these counterexamples: For a sequence of pairs of RVs $\lrc{\lrp{X_n, Y_n}}$ that converge to $(X,Y)$ in probability, \mbox{$(X_n,Y_n)\stochto{p}(X,Y)$}, either
\begin{itemize}
\item 
    the second moment (mean power) of $\Y_n$ does not converge to that of $\Y$, viz.\ $\E{\Y^2_n} \nlongrightarrow \E{\Y^2}$,
\end{itemize}
or 
\begin{itemize}
\item 
    the MMSE in estimating $\Y$ from $\X_n$ is strictly better than the MMSE of estimating $\Y$ from $\X$. 
\end{itemize}

We argue that such behavior is uncharacteristic of real-world applications and suggest adding two additional requirements:
\vspace{.2\baselineskip}
\begin{enumerate}\addtolength\itemsep{.4\baselineskip}
\item 
\label{itm:moment-convergence}
    continuity of the second moment: 
    \begin{align}
        \Limn \E{\Y^2_n} = \E{\Y^2};
    \end{align}
\item 
\label{itm:Markovity}
    a Markovian restriction 
    \begin{align}
    \label{eq:Markov:intro}
        \Y &\markov \X \markov \X_n  
    \end{align}
    for all $n$, depicted in \figref{fig:parallel}. This restriction ensures that the channel from $\Y$ to $\X_n$ is stochastically degraded with respect to the channel from $\Y$ to $\X$, so $\X_n$ is (Blackwell-)less informative about $\Y$ than $\X$.
\end{enumerate}
\vspace{.2\baselineskip}

We prove in \secref{s:main} that, under these two restrictions, the MMSE is \textit{Markov continuous} in probability. This also establishes MMSE continuity in probability for the additive-noises and machine-precision examples.
As a step in this proof, we prove that the MMSE is upper semicontinuous (u.s.c.)\ in distribution as long as the second moment of $\Y_n$ converges to that of $\Y$ (requirement~\ref{itm:moment-convergence} above).
We further establish MMSE u.s.c.\ in distribution for the case of a common parameter $\Y = \Y_n$ with finite second moment, and its continuity
if, in addition, the induced channel (conditional distribution) from $\Y$ to $\X_n$ is stochastically degraded with respect to the channel from $\Y$ to $\X$. 

In \secref{ss:linear-MMSE}, we supplement the above results by proving that the MMSE under linear estimation (LMMSE) is continuous in distribution as long as the second moments of $\seq{X_n}$ and of $\seq{Y_n}$ converge to those of $X$ and $Y$, respectively, provided the covariance matrix of $\X$ is invertible. 

In \secref{s:general-cost}, we generalize the results of \secref{s:main} from the squared Euclidean distance cost inherent to the MMSE to Bregman divergence cost functions and continuous metric cost functions. Note that the Kullback--Leibler divergence, denoted by $\mathrm{D}_\mathrm{KL}$, is an instance of a Bregman divergence. Further, the Jensen--Shannon divergence~\cite{Sibson:Jensen-Shannon-Divergence:Springer1969,Lin:Jensen-Shannon-Divergence:TIT1991}, defined as 
\begin{align}
\label{eq:Jensen--Shannon-Divergence:def}
    \begin{aligned}
    \JSD{P}{Q} \triangleq {}&\half \KL{P}{\frac{P+Q}{2}}
     + \half \KL{Q}{\frac{P+Q}{2}},
    \end{aligned}
\end{align}
is the square of a metric~\cite{Endres-Schndelin:Jensen-Shannon-Divergence:Metric:TIT2003,Osterreicher-Vajda::Jensen-Shannon-Divergence:Metric:AISM2003} and is continuous~\cite{Topsoe:Jensen-Shannon-Divergence-Pinsker-inequality:TIT2000,Briet-Harremoes:Jensen-Shannon-Divergence:continuity:PhsRevA2009} over standard simplices. Hence, our generalizations establish the Markovian continuity of the minimal mean Kullback--Leibler and Jensen--Shannon costs over standard simplices.

We establish all the results in this work in the more general framework of RVec parameters and measurements. A summary of the main results is available in \tabref{tab:main_results} in \secref{s:main}.

We conclude the paper with \secref{s:summary} by a summary and discussion of possible future directions.

Next, we introduce the notation used in this paper;
necessary background about stochastic convergence and stochastic degradedness is provided in \appref{app:basics}.

\begin{figure*}[!th]
    \resizebox{\textwidth}{!}{
\begin{tikzpicture}[auto, arrow/.style={very thick, ->, >=stealth'},start chain=going right,>=latex,node distance=2.5cm,>=latex']


    \node (X) {};
    \node[block, right = 15mm of X] (Y) {$P_{\X|\Y}$};
    \node[right = 15mm of Y] (dots1) {$\dots$};
    \node[block, right = 15mm of dots1] (Yn) {$P_{\X_{n}|\X_{n+1}}$};
    \node[right = 15mm of Yn] (dots2) {$\dots$};
    \node[block, right = 15mm of dots2] (Y2) {$P_{\X_2|\X_3}$};
    \node[block, right = 15mm of Y2] (Y1) {$P_{\X_1|\X_2}$};
    \node[right = 15mm of Y1] (Y1out) {};

    \draw[arrow] (X) -- node[above] {$\Y$} (Y);
    \draw[arrow] (Y) -- node[above] {$\X$} (dots1);
    \draw[arrow] (dots1) -- node[above] {$\X_{n+1}$} (Yn);
    \draw[arrow] (Yn) -- node[above] {$\X_{n}$} (dots2);
    \draw[arrow] (dots2) -- node[above] {$\X_{3}$} (Y2);
    \draw[arrow] (Y2) -- node[above] {$\X_{2}$} (Y1);
    \draw[arrow] (Y1) -- node[above] {$\X_{1}$} (Y1out);
\end{tikzpicture}\space
}
    \caption{Illustration of a nested sequence of (physically) degraded channels
    $\Y \markov \X \markov \cdots \markov \X_{n+1} \markov \X_n \markov \cdots \markov \X_3 \markov \X_2 \markov \X_1$.} 
\label{fig:nested-garbling}
\end{figure*}

\begin{figure*}[t]
    \resizebox{\textwidth}{!}{
\begin{tikzpicture}[auto, arrow/.style={very thick, ->, >=stealth'},start chain=going right,>=latex,node distance=2.5cm,>=latex']


    \node (X) {};
    \node[block, right = 15mm of X] (Y) {$P_{\X|\Y}$};
    \node[right = 15mm of Y] (dots1) {$\dots$};
    \node[block, right = 15mm of dots1] (Yn) {$P_{\X_{n}|\X}$};
    \node[right = 15mm of Yn] (dots2) {$\dots$};
    \node[block, right = 15mm of dots2] (Y2) {$P_{\X_2|\X}$};
    \node[block, right = 15mm of Y2] (Y1) {$P_{\X_1|\X}$};
    \node[right = 15mm of Y1] (Y1out) {};

    \node[above right = 5mm and 5mm of Y] (YorigR) {};
    \node[above = 5mm of Y] (Yorig) {};
    
    \node[above left = 5mm and 5mm of Yn] (YnL) {};
    \node[above = 5mm of Yn] (YnAbove) {};
    \node[above right = 5mm and 5mm of Yn] (YnR) {};

    \node[above left = 5mm and 5mm of Y2] (Y2L) {};

    \node[below = 5mm of Yn] (YnBelow) {};
    \node[below = 5mm of Y2] (Y2Below) {};
    \node[below = 5mm of Y1] (Y1Below) {};

    \draw[arrow] (X) -- node[above] {$\Y$} (Y);
    \draw[arrow] (Y) |- node[pos=.8] {$\X$} (YorigR); 
    \draw[very thick, dotted] (YorigR) -- (YnL);
    \draw[arrow] (YnL) -| (Yn);
    \draw[very thick] (Yn) |- (YnR);
    \draw[very thick, dotted] (YnR) -- (Y2L);
    \draw[arrow] (Y2L) -| (Y2);
    \draw[arrow] (Y2L) -| (Y1);

    \draw[arrow] (Yn) -- node[pos=1] {$\X_n$} (YnBelow);
    \draw[arrow] (Y2) -- node[pos=1] {$\X_2$} (Y2Below);
    \draw[arrow] (Y1) -- node[pos=1] {$\X_1$} (Y1Below);
\end{tikzpicture}\space
}
    \caption{Illustration of a sequence of individually (physically) degraded channels: 
    $\Y \markov \X \markov \X_i$ for all $i \in \firstnats{n}$. This is a less stringent requirement than the one depicted in \figref{fig:nested-garbling} as it does not assume degradedness between $\X_i$ and $\X_j$ for $i \neq j$.} 
    \label{fig:parallel}
\end{figure*}

\subsection{Notation}

\begin{IEEEdescription}[\IEEEusemathlabelsep\setlength{\labelwidth}{0.31\columnwidth}\setlength{\itemsep}{0.15\baselineskip}]
\item[{\parbox[t]{\labelwidth}{\raggedright $\reals, \reals_{>0}, \reals_{\geq 0}, \nats, \ints$}}] The sets of reals, positive reals, non-negative reals, positive integers, and integers, respectively.
\item[{\parbox[t]{\labelwidth}{\raggedright $x[i]$}}] The $i^\mathrm{th}$ entry of vector $x \in \reals^k$ for $i \in \firstnats{k}$.
\item[{\parbox[t]{\labelwidth}{\raggedright $x^T$}}] The transpose of a vector $x$.
\item[{\parbox[t]{\labelwidth}{\raggedright $x^r$}}] $\lrp{x^r[1], x^r[2], \ldots, x^r[k]}^T$ for $r \in \reals$, $x \in \reals^k$, and $k \in \nats$ 
(componentwise exponentiation).
\item[{\parbox[t]{\labelwidth}{\raggedright $\inner{x}{y}$}}] $x^T y$---the standard Euclidean inner product between vectors $x, y \in \reals^k$ for \mbox{$k\in \nats$}.
\item[{\parbox[t]{\labelwidth}{\raggedright $\abs{x}$}}] The componentwise absolute value of $x \in \reals^k$.
\item[{\parbox[t]{\labelwidth}{\raggedright $\norm{x}_r$}}] $\lrp{\sum_{i=1}^k \abs{x[i]}^r}^{1/r}$ for $r \in(0,\infty)$, and $\max\limits_{1\leq i \leq k} \abs{x[i]}$ for $r=\infty$, for $x \in \reals^k$.
\item[{\parbox[t]{\labelwidth}{\raggedright $\norm{x}$}}] $\norm{x}_2$---the standard Euclidean norm of a vector $x$.
\item[{\parbox[t]{\labelwidth}{\raggedright $\nabla{\phi(x)}$}}] The gradient of $\phi$ at $x \in \reals^k$.
\item[{\parbox[t]{\labelwidth}{\raggedright $x \leq y$}}] $x[i] \leq y[i]$ for all $i \in \lrc{1, 2, \ldots, k}$, where $x,y \in \reals^k$ and $k \in \nats$.
\item[{\parbox[t]{\labelwidth}{\raggedright $\floor{x}$}}] The componentwise floor operation applied on $x \in \reals^k$ for $k \in \nats$.
\item[{\parbox[t]{\labelwidth}{\raggedright $\sign\lrc{x}$}}] The sign of $x \in \reals$.
\item[{\parbox[t]{\labelwidth}{\raggedright $\trace\lrc{A}$}}] The trace of $A \in \reals^{k \times k}$ for $k \in \nats$.
\item[{\parbox[t]{\labelwidth}{\raggedright $\lim, \limsup, \liminf$}}] Limit, limit superior, and limit inferior, respectively.
\item[{\parbox[t]{\labelwidth}{\raggedright $\mathrm{E}, \mathrm{Var}$}}] Expectation and variance operators, respectively.
\item[{\parbox[t]{\labelwidth}{\raggedright $\innerE{X}{Y}$}}] $\E{\inner{X}{Y}} = \E{X^T Y}$ for random vectors (RVecs) \mbox{$X, Y \in \reals^k$} where $k \in \nats$.
\item[{\parbox[t]{\labelwidth}{\raggedright $\normEr{X}$}}] $\lrp{\E{\sum_{i=1}^k \abs{X[i]}^r}}^{1/r}$
 for an RVec $X \in \reals^k$ where $k \in \nats$ and $r \in \reals_{>0}$; for $r=2$,  $\sqrt{\innerE{X}{X}} = \sqrt{\E{X^T X}}$.
\item[{\parbox[t]{\labelwidth}{\raggedright $X \stackrel{d}= Y$}}] $X$ and $Y$ are identically distributed.
\item[{\parbox[t]{\labelwidth}{\raggedright $X \indep Y$}}] Independence between RVecs $X$ and $Y$.
\item[{\parbox[t]{\labelwidth}{\raggedright $X \markov Y \markov Z$}}] Markov triplet: $X$ and $Z$ are independent given $Y$.
\item[{\parbox[t]{\labelwidth}{\raggedright $X \dmarkov Y \dmarkov Z$}}] Garbled triplet: the conditional distribution of $Z$ given $X$ is stochastically degraded/garbled with respect to the conditional distribution of $Y$ given $X$ (see also \defref{def:garbled}).
\item[{\parbox[t]{\labelwidth}{\raggedright $X_n \stochto{d} X$}}] Convergence in distribution of $\seq{X_n}$ to~$X$.
\item[{\parbox[t]{\labelwidth}{\raggedright $X_n \stochto{p} X$}}] Convergence in probability of $\seq{X_n}$ to~$X$.
\item[{\parbox[t]{\labelwidth}{\raggedright $X_n \stochto{a.s.} X$}}] Almost-sure convergence of $\seq{X_n}$ to~$X$.
\item[{\parbox[t]{\labelwidth}{\raggedright $X_n \stochto{m.s.} X$}}] Convergence in mean square (m.s.) of $\seq{X_n}$ to~$X$.
\item[{\parbox[t]{\labelwidth}{\raggedright $\MMSE{\Y}{\X}$}}] The MMSE in estimating $\Y$ given $\X$.
\item[{\parbox[t]{\labelwidth}{\raggedright $\LMMSE{\Y}{\X}$}}] The LMMSE in linearly estimating $\Y$ given $\X$.
\end{IEEEdescription}


\section{Discussion of Existing Results}
\label{s:existing}

We first present the definition of the MMSE~\cite[Chapter~8]{Gubner:Random-Processes-for-EE:Book:2006}, \cite[Chapter~4]{Lehmann-Casella:estimation:book2006},\cite[Chapter~7]{PapoulisBook:4thEd}.

\begin{defn}[MMSE]
\label{def:MMSE}
    The MMSE in estimating an RVec $\Y$ with a finite second moment, $\normE{\Y}<\infty$,
    from an RVec $\X$ is defined as 
    \begin{align}
        \MMSE{\Y}{\X} \defeq \inf \normE{\Y - \hY}^2 ,
    \end{align}
    where the infimum is over all RVecs $\hY$ 
    with finite second moment that satisfy 
    $\Y \markov \X \markov \hY$; without loss of generality, the infimum may be restricted to $\hY = g(\X)$ for measurable functions $g$ with $\normE{\hY} < \infty$.
\end{defn}

The following is a known characterization of the MMSE \cite[Chapter~4]{Lehmann-Casella:estimation:book2006}, \cite[Chapter~9.1.5]{Intro-probability:MMSE-estimation:Book}, \cite[Appendix for Chapter 3]{Babook:Estimation} which is often used as its definition.

\begin{thm}[MMSE and MMSE estimator]
\label{thm:MMSE}
    The MMSE estimate of an RVec $\Y$ with a finite second moment, $\normE{\Y}<\infty$, from an RVec $\X$ is given by $\CE{\Y}{\X}$, and the corresponding MMSE is given as 
    \begin{subequations}
    \noeqref{eq:thm:MMSE:def}
    \label{eq:thm:MMSE}
    \begin{align}
        \MMSE{\Y}{\X} 
        &= \normE{\Y - \CE{\Y}{\X}}^2
    \label{eq:thm:MMSE:def}
    \\ &
        = \normE{\Y}^2 - \normE{\CE{\Y}{\X}}^2.
    \label{eq:thm:MMSE:diff-var}
    \end{align}
    \end{subequations}
\end{thm}

It is well known that the MMSE is not continuous in general \cite{Wu-Verdu:MMSE:IT2012,Yuksel-Linder:continuity-control:SICON2012,Mori-Szekely:4axioms-dependence:Metrika2019}, \cite[Chapter~8.3]{Yuksel-Basar:games:book:2024}.
We start by recalling known counterexamples that demonstrate it.   

We first demonstrate that, even when the measurements are independent of the parameter (or absent), the MMSE, which then reduces to the variance, might not be continuous.
\begin{example}
\label{ex:not-m.s.}
    Let $\X_n = \X$ be a standard Gaussian RV for all $n \in \nats$. Set $\Y = 0$ and 
    \begin{align}
        \Y_n = 
        \begin{cases}
            \sqrt{n}, & \mathrm{w.p.}\ \frac{1}{2n}
         \\ -\sqrt{n}, & \mathrm{w.p.}\ \frac{1}{2n}
         \\ 0, & \mathrm{w.p.}\ 1 - \frac{1}{n} ,
        \end{cases}
    \end{align}
    independent of $\X$.
    Clearly, $\lrp{\Y_n, \X_n} \stochto{d} \lrp{\Y, \X} = \lrp{0, \X}$, but 
    \begin{align}
        \Limn \normE{\Y_n - \Y} = \Limn \normE{\Y_n} = \Limn 1 = 1, 
    \end{align}
    meaning that $\Y_n \stackrel[n \to \infty]{m.s.}\nlongrightarrow \Y$. Furthermore,
    \begin{align}
        \Limn \E{\Y_n^2} = 1 > 0 = \E{\Y^2}.
    \end{align}
    Now, since $\Y_n \indep \X_n$ for all $n \in \nats$,
    \begin{align}
        \MMSE{\Y_n}{\X_n} &= \normE{\Y_n}^2 = 1 
        > 0 = \MMSE{\Y}{\X} 
    \end{align}
    meaning that the MMSE is not continuous in this case:
    \begin{align}
        \limn \MMSE{\Y_n}{\X_n} &= 1 > 0 = \MMSE{\Y}{\X}.
    \end{align}
    The latter further shows that, in this example, the MMSE is lower semicontinuous (l.s.c.) but not u.s.c.
\end{example}

Even when $\Y_n \stochto{m.s.} \Y$, the MMSE might not be continuous when a Markovian restriction of the form \eqref{eq:Markov:intro} does not hold. This is demonstrated in the following two examples.
\begin{example}
\label{ex:X+Y/n}
    Let $X$ and $Y$ be independent RVs such that
    $\Y$ has bounded support and $\X \in \ints$ with $\normE{\X} < \infty$.
    For concreteness, let $\X$ be an equiprobable Bernoulli RV, and let $\Y$ be uniformly distributed over the unit interval.
    Define $\Y_n = \Y$ and 
    \begin{align}
        \X_n = \X + \frac{\Y}{n} 
    \end{align}
    for all $n \in \nats$.

    Since $\Y_n = \Y$ for all $n \in \nats$ and $\normE{\Y} < \infty$, $\Y_n \stochto{m.s.} \Y$ and $\Y_n \stochto{a.s.} \Y$ trivially hold. 
    
    Since $\Y$ is bounded, $\normE{\Y} < \infty$. Consequently, 
    \begin{align} 
        \normE{\X_n} = \normE{\X + \frac{\Y}{n}} \leq \normE{\X} + \frac{1}{n} \normE{\Y} < \infty .
    \end{align}
    Furthermore,  
    \begin{align} 
        \Limn \normE{\X_n - \X} 
        = \Limn \normE{\frac{\Y}{n}} = 0, 
    \end{align}
    Hence, $\X_n \stochto{m.s.} \X$.
    Furthermore, $\seq{\X_n} \stochto{a.s.} \X$.

    \vspace{.5\baselineskip}
    Since $X \indep Y$, $\MMSE{\Y}{\X} = \Var{\Y} = 1/12$.
    
    However, since $\Y_n = \Y$ can be perfectly estimated from~$\X_n$, \viz\ $\Y = n \lrp{\X_n - \floor{\X_n}}$ a.s., 
    \begin{align}
        \MMSE{\Y_n}{\X_n} &= \MMSE{\Y}{\X_n} = 0 &\forall n \in \nats.
    \end{align}
    
    Hence, the MMSE is not continuous in this example:
    \begin{align}
        \Limn \MMSE{\Y_n}{\X_n} = 0 < \frac{1}{12} = \MMSE{\Y}{\X}.
    \end{align}
    In particular, the MMSE is u.s.c.\ but not l.s.c.\ in this example.
    Note further that the Markov relation \eqref{eq:Markov:intro} does not hold in this example.
\end{example}

\begin{example}
\label{ex:Yn+N}
    Let $\Y$ and $N$ be independent Rademacher RVs, 
    and $\X = \Y + N$. In particular, $\Var{\Y} = 1 < \infty$.
    Let $\Y_n = \frac{n}{n+1} \Y$ and $\X_n = \Y_n + N$ for all $n \in \nats$.

    Clearly, $\lrp{X_n, Y_n} \stochto{a.s.} \lrp{X,Y}$ and 
    \begin{align}
        \Limn \E{\Y_n^2} &= \Limn \lrp{\frac{n}{n+1}}^2 \E{\Y^2} = \E{\Y^2} ,
     \\ \Limn \E{\X_n^2} &= \Limn \E{\Y_n^2} + \E{N^2} 
     \\ &= \E{\Y^2} + \E{N^2} = \E{\X^2}.
    \end{align}
    Hence 
    $\lrp{X_n, Y_n} \stochto{m.s.} \lrp{X,Y}$ (see Theorems~\ref{lem:convergence:subseq<=>seq} and \ref{thm:Vitali:prob+2ndMom-->m.s.}).

    Since $\Y_n = \X_n - \sign\lrc{\X_n}$ a.s., $\Y_n$ can be perfectly estimated from $\X_n$. Hence, $\MMSE{\Y_n}{\X_n} = 0$ for all $n \in \nats$.
    However, $\Y$ cannot be perfectly estimated from $\X = \Y + N$. In fact, $\CE{\Y}{\X} = \X/2$ and 
    \begin{align}
        \MMSE{\Y}{\X} = \E{\lrp{\frac{\Y - N}{2}}^2} = \half.
    \end{align}
    Hence, the MMSE is not continuous:
    \begin{align}
        \Limn \MMSE{\Y_n}{\X_n} = 0 < 1/2 = \MMSE{\Y}{\X}. 
    \end{align}
    Again, the latter shows that the MMSE is u.s.c.\ but not l.s.c.\ in this example. 
    And again, we note that the Markov relation \eqref{eq:Markov:intro} does not hold in this example.
\end{example}

While MMSE is not generally continuous or even semicontinuous, 
it was proved by Wu and Verd\'u~\cite[Theorem~3]{Wu-Verdu:MMSE:IT2012} 
to be u.s.c.\ if the supports of the RVecs $\Y$ and $\seq{\Y_n}$ are uniformly bounded (a.s.\ bounded by the same constant). 

\begin{thm}[\!\!{\cite[Theorem~3]{Wu-Verdu:MMSE:IT2012}}]
\label{thm:Wu-Verdu:usc}
    Let $(X,Y)$ be a pair of RVecs and let $\seq{\lrp{X_n, Y_n}}$ be a sequence of pairs of RVecs such that 
    \begin{itemize}\addtolength\itemsep{.2\baselineskip}
    \item
        $\lrp{X_n,Y_n} \stochto{d} (X,Y)$;
    \item
        $\PR{\norm{\Y} \leq m} = 1$ and $\PR{\norm{\Y_n} \leq m} = 1$ for all $n \in \nats$ for some $m \in \reals$.
    \end{itemize}
     Then, the MMSE is u.s.c.\ in distribution:
     \begin{align}
     \label{eq:thm:Wu-Verdu:usc}
         \limsup_{n \to \infty} \MMSE{\Y_n}{\X_n} \leq \MMSE{\Y}{\X}.
     \end{align}
\end{thm}

When restricting the possible statistical relations, the following continuity results have been proved.

\begin{thm}[\!\!{\cite[Theorem~4]{Wu-Verdu:MMSE:IT2012}}]
\label{thm:Wu-Verdu:additive-noise}
    Let $X,Y,$ and $N$ be RVecs of the same length. Let $\seq{\lrp{X_n, Y_n}}$ be a sequence of pairs of RVecs, such that 
    \begin{itemize}\addtolength\itemsep{.2\baselineskip}
    \item 
        $\X = \Y + N$, and $\X_n = \Y_n + N$ for all $n \in \nats$;
    \item 
        $N \indep \Y, \seq{\Y_n}$;
    \item 
        $\normE{\Y}, \normE{N} < \infty$;
    \item 
        $\Y_n \stochto{d} \Y$.
    \end{itemize}
    Then, 
    \begin{itemize}\addtolength\itemsep{.2\baselineskip}
    \item 
         The MMSE is u.s.c.~\eqref{eq:thm:Wu-Verdu:usc} in distribution.\footnote{In \thmref{thm:Wu-Verdu:additive-noise} and \remref{rem:Wu-Verdu:MMSE(Xn|Yn)=MMSE(N)(Yn)}, MMSE denotes the infimum over all measurable estimators, without requiring the estimand or estimator to have a finite second moment.}
    \item 
        In addition, if $N$ has a PDF that is bounded and continuous, then the MMSE is continuous in distribution:
        \begin{align}
            \limn \MMSE{\Y_n}{\X_n} = \MMSE{\Y}{\X}.
        \end{align}
    \end{itemize}
\end{thm}

Unfortunately, the result above is limited to additive noise channels where the noise $N$ has a bounded and continuous PDF.
In particular, it does not guarantee MMSE continuity, e.g., for noises with continuous uniform or arcsine distributions, 
or noises whose distribution contains discrete or singular behavior (recall Lebesgue's decomposition theorem~\cite[Chapter~2, Section~2.3]{Gut:Probability:Book2005}).

\begin{remark}
\label{rem:Wu-Verdu:MMSE(Xn|Yn)=MMSE(N)(Yn)}
    Furthermore, as indicated in \cite{Wu-Verdu:MMSE:IT2012}, since 
    \begin{align}
        \MMSE{\Y_n}{\X_n} 
        &= \inf_{g_1} \normE{\Y_n - g_1\lrp{\X_n}}^2
     \\ &= \inf_{g_1} \normE{\Y_n - \X_n + \X_n - g_1\lrp{\X_n}}^2
     \\ &= \inf_{g_2} \normE{N - g_2\lrp{\X_n}}^2
     \\ &= \MMSE{N}{\X_n} ,
    \end{align}
    the setting of \thmref{thm:Wu-Verdu:additive-noise} can be viewed as an estimation problem of a fixed parameter $N$ from $\X_n$, where $\X_n$ is the output of an additive noise channel with noise $\Y_n$, and where $\X_n \stochto{d} \X$. Since $N \indep \Y, \seq{\Y_n}$, this means further that 
    \begin{align}
        \lrp{N, \Y_n, \X_n} \stochto{d} \lrp{N, \Y, \X}.
    \end{align}
\end{remark}

The framework where a common parameter $\Y$ passes through a sequence of channels resulting in a sequence $\seq{\X_n}$ was also studied by Y\"uksel and Linder \cite[Theorem~3.2]{Yuksel-Linder:continuity-control:SICON2012} and Hogeboom-Burr and Y\"uksel~\cite{Hogeboom-Yuksel:continuity:SICON2023,Hogeboom_thesis}
(see also \cite[Chapter~8.3]{Yuksel-Basar:games:book:2024}).
The following theorem and remarks summarize relevant results about  continuity in distribution.

\begin{thm}[{\cite{Hogeboom_thesis}, \cite[Theorem~8.3.4]{Yuksel-Basar:games:book:2024}}]
\label{thm:Yuksel:de-garbled:convergence}
    Let $c: \cY \times \mathcal{\hY} \to \reals$, $\Y \in \cY$ be an RVec, and $\seq{\X_n \in \cX}$ be a sequence of RVecs such that
    \begin{enumerate}\addtolength\itemsep{.2\baselineskip}
    \item 
    \label{itm:Yuksel:de-garbled:convergence:bounded-func}
        $c$ is continuous and bounded;
    \item 
    \label{itm:Yuksel:de-garbled:convergence:convex-estimate}
        $\mathcal{\hY}$ is a convex set;
    \item 
    \label{itm:Yuksel:de-garbled:convergence:weak-convergence}
        $\lrp{\Y, \X_n} \stochto{d} \lrp{\Y, \X}$;
    \item 
    \label{itm:Yuksel:de-garbled:convergence:X--Y--Y[n]}
        $\Y \dmarkov \X \dmarkov \X_n$ for all $n \in \nats$;
    \item 
    \label{itm:Yuksel:de-garbled:convergence:X--Y[n+1]--Y[n]}
        $\Y \dmarkov \X_{n+1} \dmarkov \X_n$ for all $n \in \nats$.
    \end{enumerate}
    Then, the minimum mean cost is continuous in distribution:
    \begin{align}
        \Limn \inf_{g:\ \cX \to \mathcal{\hY}} \E{c \lrp{\Y, g\lrp{\X_n}}}
        = \inf_{g:\ \cX \to \mathcal{\hY}} \E{c \lrp{\Y, g\lrp{\X}}} .
    \end{align}
\end{thm}

Requirement~\eqref{itm:Yuksel:de-garbled:convergence:X--Y--Y[n]}, $\Y \dmarkov \X \dmarkov \X_n$, means that the channel from $\Y$ to $\X_n$ is stochastically degraded with respect to the channel from $\Y$ to $\X$. 
Equivalently, this requirement means that there exist probabilistically identical channels to these two channels such that,
for the same input $\Y$, their outputs satisfy~\eqref{eq:Markov:intro}. 
See~\appref{app:basics} for further details about stochastic degradedness.

\begin{remark}
\label{rem:Yuksel:bounded-RVs}
    For uniformly bounded RVecs $\Y$, $\X$, and $\seq{\X_n}$, \thmref{thm:Yuksel:de-garbled:convergence} can be readily applied to the MMSE by selecting the cost function to be the squared Euclidean distance.
\end{remark}

\begin{remark}
\label{rem:Yuksel:usc}
    When only requirements~\ref{itm:Yuksel:de-garbled:convergence:bounded-func}--\ref{itm:Yuksel:de-garbled:convergence:weak-convergence} hold, 
    Y\"uksel and Linder \cite[Theorem~3.2]{Yuksel-Linder:continuity-control:SICON2012} (see also \cite[Theorem~8.3.3]{Yuksel-Basar:games:book:2024}) proved that u.s.c.\ in distribution holds:
    \begin{align}
        \limsup_{n \to \infty} \inf_{g:\ \cX \to \mathcal{\hY}} \E{c \lrp{\Y, g\lrp{\X_n}}}
        \leq \inf_{g:\ \cX \to \mathcal{\hY}} \E{c \lrp{\Y, g\lrp{\X}}} .
    \end{align}
    This is subsumed by the result of \thmref{thm:Wu-Verdu:usc} for a squared Euclidean distance $c$ and bounded RVecs that take values in Euclidean spaces; see also \secref{s:general-cost}.
\end{remark}

While \thmref{thm:Yuksel:de-garbled:convergence} extends the continuity guarantees beyond the scope of additive noise channels of \thmref{thm:Wu-Verdu:additive-noise}, it is limited to bounded costs and \textit{nested} garbling of $\seq{\X_n}$ and $\X$ (see also \figref{fig:nested-garbling}):
\begin{align}
\label{eq:nested-garbling}
    \Y \dmarkov \X \dmarkov \cdots \dmarkov \X_{n+1} \dmarkov \cdots \dmarkov \X_2 \dmarkov \X_1.
\end{align}
This is demonstrated by the following additive-noise example.

\begin{example}
\label{ex:Yn=X+N/n}
    Let $\Y$ and $N$ be independent continuous RVs uniformly distributed over the interval $[-\sqrt{3},\sqrt{3}]$. Let $\X_n = \Y + {N}/\lrp{n+1}$ and $\X = \Y$.
    Note that the second moments are all finite: $\normE{\Y} = \normE{\X} = 1$ and $\E{\X_n^2} = 1 + 1/(n+1)^2 \leq 2$ for all $n \in \nats$. Furthermore, 
    \begin{align}
        \limn \normE{\X_n - \X} = \limn \frac{\normE{N}}{n+1} = 0.
    \end{align}
    Therefore, $\X_n \stochto{m.s.} \X = \Y$, meaning that
    \begin{align}
        \Limn \MMSE{\Y}{\X_n} = 0 = \MMSE{\Y}{\X}
    \end{align}    
    by the squeeze theorem:
    \begin{align}
        0 \leq \limn \MMSE{\Y}{\X_n} \leq \limn \normE{\Y - \X_n}^2 = 0.
    \end{align}
    However, since requirement~\ref{itm:Yuksel:de-garbled:convergence:X--Y[n+1]--Y[n]} in \thmref{thm:Yuksel:de-garbled:convergence} does not hold for any $n \in \nats$, this theorem cannot be applied for this case.\footnote{Physical degradedness $\Y \markov \X_{n+1} \markov \X_{n}$ clearly does not hold. One can verify that stochastic degradedness cannot hold either.} The conditions of \thmref{thm:Wu-Verdu:additive-noise} (recall \remref{rem:Wu-Verdu:MMSE(Xn|Yn)=MMSE(N)(Yn)}) do not hold either since the PDF of the uniform distribution is not continuous at the support boundaries.
\end{example}

\begin{remark}
    The failure of requirement~\eqref{eq:nested-garbling} in \exref{ex:Yn=X+N/n}
    stems from the rate at which the noise vanishes rather than from its distribution.
    Let $X$ and $N$ be as in \exref{ex:Yn=X+N/n}, and let 
    $Y_n = X + N/\rho^n$ 
    for some $\rho > 1$.
    If $\rho \geq 2$ is an integer, $\seq{Y_n}$ satisfies~\eqref{eq:nested-garbling}.
    Indeed, for any $s > 0$, a uniform RV over $[-\rho s, \rho s]$ is distributed as the sum
    of a uniform RV over $[-s, s]$ and an independent RV that is uniform over
    $\{-(\rho-1)s, -(\rho-3)s, \ldots, (\rho-1)s\}$. Taking $s = \sqrt{3}/\rho^{n+1}$ shows that the channel to $Y_n$ can be obtained from the channel to $Y_{n+1}$ by adding independent noise, and hence
    \thmref{thm:Yuksel:de-garbled:convergence} applies.
    If $1 < \rho < 2$, requirement~\eqref{eq:nested-garbling} does not hold, as in
    \exref{ex:Yn=X+N/n}.
    Nevertheless,
    $\lim_{n \to \infty} \mathrm{MMSE}\left(X \middle| Y_n\right) = 0 = \mathrm{MMSE}\left(X \middle| Y\right)$
    for every $\rho > 1$, by the same squeeze argument.
    Thus, whether \thmref{thm:Yuksel:de-garbled:convergence} applies hinges on a feature that is immaterial to the continuity of the MMSE.
    \thmref{thm:garbled:weak-continuity} in \secref{s:main} dispenses with~\eqref{eq:nested-garbling} and covers all these cases.
\end{remark}

While these results provide guarantees for the continuity of the MMSE for certain cases, their scope remains limited. 
In \secref{s:main}, 
we provide guarantees for the continuity of the MMSE under a larger framework. We further supplement these results by establishing continuity in distribution of the MMSE under linear estimation in \secref{ss:linear-MMSE}.


\section{Main Results} 
\label{s:main}

\begin{table*}[t]
\centering
\small
\renewcommand{\arraystretch}{1.15}
\setlength{\tabcolsep}{6pt}
\caption{Summary of continuity guarantees for MMSE / related costs.}
\label{tab:main_results}
\begin{tabularx}{\textwidth}{@{}>{\raggedright\arraybackslash}p{1.3cm}
                                >{\raggedright\arraybackslash}p{2.6cm}
                                >{\raggedright\arraybackslash}p{4.9cm}
                                >{\raggedright\arraybackslash}p{3.5cm}
                                >{\raggedright\arraybackslash}X@{}}
\toprule
\toprule
\textbf{Result} &
\textbf{Mode of convergence} &
\textbf{Moment condition} &
\textbf{Additional constraint} &
\textbf{Conclusion} \\
\midrule
\midrule

Thm.~\ref{thm:MMSE-usc} &
$(X_n,Y_n)\xrightarrow{d}(X,Y)$ &
$\Lim_{n\to\infty} \E{X_n^2} = \E{X^2}$ &
\centering --- &
MMSE is u.s.c. \\
\midrule

Thm.~\ref{thm:garbled:weak-continuity} &
$(X,Y_n)\xrightarrow{d}(X,Y)$ (common $X$) &
$\|X\|_{L^2}<\infty$ &
$X \dmarkov Y \dmarkov Y_n$ (stochastic degradedness / garbling) &
MMSE is continuous
\\
\midrule

Thm.~\ref{thm:MMSE:Markov-continuity} &
$(X_n,Y_n)\xrightarrow{p}(X,Y)$ &
$\Lim_{n\to\infty}\E{X_n^2} = \E{X^2}$ &
$X \markov Y \markov Y_n$ (physical degradedness) &
MMSE is continuous
\\
\midrule

Thm.~\ref{thm:Markov-continuity:estimator} &
$(X_n,Y_n)\xrightarrow{p}(X,Y)$ &
$\Lim_{n\to\infty}\E{\abs{X_n}^r} = \E{\abs{X}^r}$ for $r \geq 1$&
$X \markov Y \markov Y_n$ (physical degradedness) &
$\mathbb{E}[X_n|Y_n]\xrightarrow{L^r}\mathbb{E}[X|Y]$
\\
\midrule

Thm.~\ref{thm:LMMSE:continuity} &
$(X_n,Y_n)\xrightarrow{d}(X,Y)$ &
$\E{X_n^2} \to \E{X^2}$, $\E{Y_n^2} \to \E{Y^2}$ &
$C_Y$ is invertible &
LMMSE is continuous 
\\
\midrule

Thm.~\ref{thm:Bregman-cost:Markov-continuity} &
$(X_n,Y_n)\xrightarrow{p}(X,Y)$ &
$\Lim_{n\to\infty}\E{\abs{X_n}} = \E{|X|}$, $\Limn \E{\abs{\phi(X_n)}} = \E{\abs{\phi(X)}}$ &
$X \markov Y \markov Y_n$ (physical degradedness) &
Bregman divergence associated with $\phi$ is continuous
\\
\midrule

Thm.~\ref{thm:gen-cost:usc:metric-cost} &
$(X_n,Y_n)\xrightarrow{d}(X,Y)$ &
$\E{\cost{\Y_n}{\oy}} \to \E{\cost{\Y}{\oy}}$ &
$c$ is a continuous metric cost &
Minimal mean cost is u.s.c. \\
\midrule

Thm.~\ref{thm:Markov-continuity:general-cost} &
$(X_n,Y_n)\xrightarrow{p}(X,Y)$ &
$\E{\cost{\Y_n}{\oy}} \to \E{\cost{\Y}{\oy}}$ &
$c$ is a continuous metric cost, 
$X \markov Y \markov Y_n$ (physical degradedness) &
Minimal mean cost is continuous
\\

\bottomrule
\bottomrule
\end{tabularx}
\end{table*}

In this section, we present the main results of this work, which are summarized (along with later results on more general costs) in \tabref{tab:main_results}.

We first present a result about the semicontinuity in distribution of the MMSE, which replaces the bounded-support requirement of \thmref{thm:Wu-Verdu:usc} by a relaxed requirement of continuity of the second moment. 
\begin{thm}[MMSE u.s.c.~in distribution]
\label{thm:MMSE-usc}
    Let $(X,Y)$ be a pair of RVecs and let $\seq{\lrp{X_n, Y_n}}$ be a sequence of pairs of RVecs such that 
    \begin{itemize}\addtolength\itemsep{.4\baselineskip}
    \item
        $\lrp{X_n,Y_n} \stochto{d} (X,Y)$;
    \item 
        $\Limn \E{\Y_n^2} = \E{\Y^2}$ and $\normE{\Y} < \infty$.\footnote{\label{foot:entrywise-squaring} Recall that the squaring in $\E{\Y^2}$ is understood componentwise.}
    \end{itemize}
     Then, the MMSE is u.s.c.\ in distribution:
     \begin{align}
     \label{eq:thm:MMSE-usc}
         \limsup_{n \to \infty} \MMSE{\Y_n}{\X_n} \leq \MMSE{\Y}{\X}.
     \end{align}
\end{thm}
The proof of \thmref{thm:MMSE-usc} is available in \appref{app:thm:MMSE-usc}.

As we have seen in Examples~\ref{ex:X+Y/n} and \ref{ex:Yn+N}, continuity of the MMSE, namely equality in \eqref{eq:thm:MMSE-usc}, does not hold in general even under the conditions of \thmref{thm:MMSE-usc}.

To establish continuity, we consider next several sets of additional requirements.

In practical scenarios, the deviation of $\X_n$ from $\X$ does not carry extra information about the nominal parameter $\Y$ beyond the information provided by the measurement $\X$. 
More precisely, a Markov relation $\Y \markov \X \markov \X_n$ (or  \mbox{$\Y \dmarkov \X \dmarkov \X_n$}), as in~\eqref{eq:Markov:intro}, holds for all $n \in \nats$.
This Markov relation, which is depicted in \figref{fig:parallel}, excludes Examples~\ref{ex:X+Y/n} and \ref{ex:Yn+N} but holds for \exref{ex:Yn=X+N/n}.

We therefore establish next continuity of the MMSE for the case of a common parameter under the degradedness condition~\ref{itm:Yuksel:de-garbled:convergence:X--Y--Y[n]}  of \thmref{thm:Yuksel:de-garbled:convergence}, $\Y \dmarkov \X \dmarkov \X_n$ by relying on Blackwell's informativeness theorem (its specialization for MMSE is stated and proved in~\lemref{lem:Blackwell}). 
Note that requirement~\ref{itm:Yuksel:de-garbled:convergence:X--Y[n+1]--Y[n]} of \thmref{thm:Yuksel:de-garbled:convergence}, $\Y \dmarkov \X_{n+1} \dmarkov \X_n$, 
is not needed; in fact, the proof of \thmref{thm:Yuksel:de-garbled:convergence} in \cite{Hogeboom_thesis}, \cite[Chapter~8.3.1]{Yuksel-Basar:games:book:2024} does not require the latter requirement. 
Since we focus on the MMSE, the cost function is taken to be the squared Euclidean distance: $c \lrp{x, \hx} = \norm{x - \hx}^2$ and is in particular unbounded. The requirement of bounded RVecs of \remref{rem:Yuksel:bounded-RVs} is also relaxed to a second-moment continuity requirement (see Theorems~\ref{thm:u.i.:properties} and \ref{thm:Vitali:prob+2ndMom-->m.s.} for more details).

\begin{thm}[MMSE continuity in distribution over degraded channels]
\label{thm:garbled:weak-continuity}
    Let $(X,Y)$ be a pair of RVecs and let $\seq{\X_n}$ be a sequence of RVecs such that 
    \begin{enumerate}\addtolength\itemsep{.5\baselineskip}
    \item
        $\lrp{\Y,\X_n} \stochto{d} (\Y,\X)$;
    \item
        $\Y \dmarkov \X \dmarkov \X_n$ for all $n \in \nats$;
    \item 
        $\normE{\Y} < \infty$.
    \end{enumerate}
    Then, the MMSE is continuous in distribution:
    \begin{align}
        \limn \MMSE{\Y}{\X_n} = \MMSE{\Y}{\X}.
    \end{align}
\end{thm}

\begin{IEEEproof}
    Since $\normE{\Y} < \infty$, the second moment of the fixed sequence $\seq{\Y}$ trivially converges to that of $\Y$. Hence, we can apply \thmref{thm:MMSE-usc} with \mbox{$\Y_n = \Y$} to attain 
    \begin{align}
    \label{eq:garbled:weak-continuity:limsup-LB}
        \limsup_{n \to \infty} \MMSE{\Y}{\X_n} \leq \MMSE{\Y}{\X}.
    \end{align}

    Now, since 
    $\Y \dmarkov \X \dmarkov \X_n$ for all $n \in \nats$, 
    by Blackwell's informativeness theorem:
    \begin{align}
        \MMSE{\Y}{\X} \leq \MMSE{\Y}{\X_n}
    \end{align}
    for all $n \in \nats$ (see \lemref{lem:Blackwell} for details). Consequently,
    \begin{align}
    \label{eq:garbled:weak-continuity:liminf-UB}
        \MMSE{\Y}{\X} \leq \liminf_{n \to \infty} \MMSE{\Y}{\X_n}.
    \end{align}

    Combining \eqref{eq:garbled:weak-continuity:limsup-LB} and \eqref{eq:garbled:weak-continuity:liminf-UB} proves the desired result.
\end{IEEEproof}

Since the random parameter is the same for all $n$, we may regard the setting of \thmref{thm:garbled:weak-continuity} as that of estimating a (common) parameter $X$ from the outputs of a converging sequence of channels  $\lrp{p_n(y|x)}_{n=1}^\infty$ converging to a channel $p(y|x)$ where $p_n$ is stochastically degraded with respect to $p$.

To dispense with the common-parameter assumption, we need  $\Y$, $\X$, and $\X_n$ to be defined on the same probability space. Hence, we elevate the convergence in distribution of $\seq{\lrp{X_n, Y_n}}$ to $\lrp{X, Y}$, to convergence in probability. This yields the following theorem.

\begin{thm}[MMSE Markovian continuity in probability]
\label{thm:MMSE:Markov-continuity}
    Let $(X,Y)$ be a pair of RVecs and let $\seq{\lrp{X_n, Y_n}}$ be a sequence of pairs of RVecs such that 
    \begin{enumerate}\addtolength\itemsep{.4\baselineskip}
    \item
    \label{itm:MMSE:Markov-continuity:Xn-->Yn}
        $\lrp{X_n,Y_n} \stochto{p} (X,Y)$;
    \item 
    \label{itm:MMSE:Markov-continuity:moment-convergence}
        $\Limn \E{\Y_n^2} = \E{\Y^2} < \infty$;\footref{foot:entrywise-squaring}
    \item 
    \label{itm:MMSE:Markov-continuity:Markovity}
        $\Y \markov \X \markov \X_n$ for all $n \in \nats$.
    \end{enumerate}
     Then, the MMSE is Markov continuous in probability:
     \begin{align}
     \label{eq:MMSE:Markov-continuity}
         \limn \MMSE{\Y_n}{\X_n} = \MMSE{\Y}{\X}.
     \end{align}
\end{thm}

\begin{IEEEproof}
    Since $\lrp{X_n,Y_n} \stochto{p} (X,Y)$, $\lrp{X_n,Y_n} \stochto{d} (X,Y)$ also holds (see \lemref{lem:convergence:a.s,m.s=>p=>d}). Hence, the MMSE is u.s.c.~\eqref{eq:thm:MMSE-usc}.
    Therefore, to establish continuity, we next prove that the MMSE is l.s.c.\ in probability under the Markov condition $\Y \markov \X \markov \X_n$.

    Set some $\eps > 0$. 
    By assumptions~\ref{itm:MMSE:Markov-continuity:Xn-->Yn} and~\ref{itm:MMSE:Markov-continuity:moment-convergence} of the theorem and 
    Vitali's convergence theorem (\thmref{thm:Vitali:prob+2ndMom-->m.s.}), 
    there exists $n_0 \in \nats$, such that 
    \begin{align}
    \label{proof:MMSE:Markov-continuity:Xn-->X}
        \normE{\Y - \Y_n} < \eps
    \end{align}
    for all $n > n_0$.
    Then, for all $n > n_0$, 
    \begin{subequations}
    \label{eq:garbled:weak-continuity:proof}
    \begin{align}
    \!\!\!
        \sqrt{\MMSE{\Y_n}{\X_n}} &= \inf_g \normE{\Y_n - g\lrp{\X_n}}
    \label{eq:garbled:weak-continuity:proof:MMSE-def1}
     \\ &\geq \inf_g \normE{\Y - g\lrp{\X_n}} - \normE{\Y_n - \Y} \quad\ \hspace{4pt}
    \label{eq:garbled:weak-continuity:proof:triangle-inequality}
     \\ & > \inf_g \normE{\Y - g\lrp{\X_n}} - \eps
    \label{eq:garbled:weak-continuity:proof:Xn-->X}
     \\ & = \sqrt{\MMSE{\Y}{\X_n}} - \eps
    \label{eq:garbled:weak-continuity:proof:MMSE-def2}
     \\ & \geq \sqrt{\MMSE{\Y}{\X}} \, -\, \eps
    \label{eq:garbled:weak-continuity:proof:Blackwell}
    ,
    \end{align}
    \end{subequations}
    where the infima are over all measurable functions with finite second moments,
    \eqref{eq:garbled:weak-continuity:proof:MMSE-def1} and \eqref{eq:garbled:weak-continuity:proof:MMSE-def2} hold by the definition of the MMSE (\defref{def:MMSE}),
    \eqref{eq:garbled:weak-continuity:proof:triangle-inequality} follows from the triangle (Minkowski) inequality~\cite[Chapter~3, Theorem~2.6]{Gut:Probability:Book2005},
    \eqref{eq:garbled:weak-continuity:proof:Xn-->X} follows from \eqref{proof:MMSE:Markov-continuity:Xn-->X},
    and 
    \eqref{eq:garbled:weak-continuity:proof:Blackwell} follows from the Markov relation $\Y \markov \X \markov \X_n$ and Blackwell's informativeness theorem (see \lemref{lem:Blackwell}).
    Since $\eps > 0$ is arbitrary, this proves the l.s.c.\ of the MMSE and completes the proof of its continuity.
\end{IEEEproof}

While \thmref{thm:MMSE:Markov-continuity} establishes the continuity of the MMSE in probability under the Markov restriction $\Y \markov \X \markov \X_n$, it does not guarantee such continuity of the MMSE estimator itself. We next prove that in fact the conditional expectation is Markov continuous in $L^r$ for $r \geq 1$.

\begin{thm}[Markovian continuity of conditional expectation in $L^r$]
\label{thm:Markov-continuity:estimator}
    Let $r \geq 1$, let $(X,Y)$ be a pair of RVecs, and let $\seq{\lrp{X_n, Y_n}}$ be a sequence of pairs of RVecs such that
    \begin{enumerate}\addtolength\itemsep{.4\baselineskip}
    \item
        $\lrp{X_n,Y_n} \stochto{p} (X,Y)$;
    \item 
    \label{itm:MMSE:Markov-continuity:estimator:moment-convergence}
        $\Limn \E{\abs{\Y_n}^r} = \E{\abs{\Y}^r}$ and $\normEr{\Y} < \infty$;\footnote{\label{foot:drop-abs}Recall that the exponentiation in $\E{\abs{\Y}^r}$ is understood componentwise.}
    \item 
        $\Y \markov \X \markov \X_n$ for all $n \in \nats$.
    \end{enumerate}
    \vspace{.5\baselineskip}
    Then,
        $\CE{\Y_n}{\X_n} \stochto{L^r} \CE{\Y}{\X}$.
\end{thm}

\begin{remark}
    The requirement $\E{\Y_n^2} \to \E{\Y^2}$  
    in Theorems~\ref{thm:MMSE-usc} and~\ref{thm:MMSE:Markov-continuity}
    can be replaced by $\Y_n \stochto{m.s.} \Y$ or by $\seq{\Y_n^2}$ being uniformly integrable (u.i.). Similarly, the requirement $\E{\abs{\Y_n}^r} \to \E{\abs{\Y}^r}$ in~\ref{thm:Markov-continuity:estimator} can be replaced by $\Y_n \stochto{L^r} \Y$ or by $\seq{\abs{\Y_n}^r}$ being uniformly integrable (u.i.) and can be stated without absolute values in certain situations. See \appref{app:basics} for details.
\end{remark}

\begin{remark}
    When $\X$ is deterministic, the Markov condition~\eqref{eq:Markov:intro} 
    reduces to $\X_n \indep \Y$ for all $n \in \nats$. This is the case in \exref{ex:X+Y/n} with $\X = 0$.
\end{remark}

To prove \thmref{thm:Markov-continuity:estimator} we use the following two lemmata.

\begin{lem}
\label{lem:E[Yn|Xn]-->E[Y|Xn]}
    Let $r \geq 1$, let $\seq{\lrp{X_n, Y_n}}$ be a sequence of pairs of RVecs, and let $\Y$ be an RVec such that 
    $\normEr{\Y} < \infty$ and 
    \mbox{$\Y_n \stochto{L^r} \Y$}. 
    Then, 
    \begin{align}
    \label{eq:E[Yn|Xn]-->E[Y|Xn]}
        \limn \normEr{\CE{\Y}{\X_n} - \CE{\Y_n}{\X_n}} = 0.
    \end{align}
\end{lem}


\begin{IEEEproof}
    Denote $\cE \triangleq \Y - \Y_n$. Then,
        $\normEr{\CE{\cE}{\X_n}}
        \leq \normEr{\cE}$, 
    by the (conditional) Jensen inequality~\cite[Chapter~9.7]{Williams:Martingales:Book1991} and the convexity of the norm function.
    Hence \eqref{eq:E[Yn|Xn]-->E[Y|Xn]} holds since $\Y_n \stochto{L^r} \Y$.
\end{IEEEproof}

\begin{lem}
\label{lem:E[Y|Xn]-->E[Y|X]}
    Let $r \geq 1$, and let $(X,Y)$ be a pair of RVecs and let $\seq{\X_n}$ be a sequence of RVecs such that 
    \begin{enumerate}\addtolength\itemsep{.5\baselineskip}
    \item
        $\X_n \stochto{p} \X$;
    \item
        $\Y \markov \X \markov \X_n$ for all $n \in \nats$;
    \item 
        $\normEr{\Y} < \infty$.
    \end{enumerate}
    Then, 
        $\CE{\Y}{\X_n} \stochto{L^r} \CE{\Y}{\X}$.
\end{lem}

The proof of \lemref{lem:E[Y|Xn]-->E[Y|X]} is available in \appref{app:proof:lem:E[Y|Xn]-->E[Y|X]}.

We are now ready to prove \thmref{thm:Markov-continuity:estimator}. 
\begin{IEEEproof}[Proof of \thmref{thm:Markov-continuity:estimator}]    
    Let $\eps > 0$, however small. 
    Since $\Limn \E{\abs{\Y_n}^r} = \E{\abs{\Y}^r}$, $\normEr{\Y} < \infty$, and $\Y_n \stochto{p} \Y$, by Vitali's convergence theorem (\thmref{thm:Vitali:prob+2ndMom-->m.s.})
    $\Y_n \stochto{L^r} \Y$.
    Then, by lemmata~\ref{lem:E[Yn|Xn]-->E[Y|Xn]} and \ref{lem:E[Y|Xn]-->E[Y|X]}, there exists $n_0 \in \nats$, such that, for all $n > n_0$, 
    \begin{align}
    \label{eq:proof:MMSE-continuity:lemmata}
    \begin{aligned}
        \normEr{\CE{\Y}{\X_n} - \CE{\Y_n}{\X_n}} &< \eps ,
     \\ \normEr{\CE{\Y}{\X} - \CE{\Y}{\X_n}} &< \eps .
    \end{aligned}
    \end{align}
    Hence, by the triangle (Minkowski) inequality, 
    \begin{align} 
        \normEr{\CE{\Y}{\X} - \CE{\Y_n}{\X_n}}
        &\leq
        \normEr{\CE{\Y}{\X_n} - \CE{\Y_n}{\X_n}}
     \\ &\qquad + \normEr{\CE{\Y}{\X} - \CE{\Y}{\X_n}}
    \nonumber
     \\ &< 2\eps.    
    \end{align}
    Since $\eps > 0$ is arbitrary, 
        $\CE{\Y_n}{\X_n} \stochto{L^r} \CE{\Y}{\X}$.
\end{IEEEproof}

Sequences that comply with the Markov restriction 
and the continuity of the second moment restriction of Theorems~\ref{thm:MMSE:Markov-continuity} and~\ref{thm:Markov-continuity:estimator} 
include corruption by independent additive noises of decreasing strength and floating point representations with increasing machine precision.   
This is summarized in the following two corollaries, which are proved in \appref{app:noise-extremes}. 
\begin{cor}[Additive noise effect]
\label{cor:additive-noises}
    Let $X,Y,M,N$ be RVecs such that \mbox{$\normE{\Y}, \normE{N} < \infty$}; $M \indep (X,Y)$; $\Y, N \in \reals^k$ for $k \in \nats$; and $\X, M \in \reals^m$ for $m \in \nats$. 
    Then, 
    \begin{align}
    \label{eq:continuity:weak_noise}
        \lim_{\lrp{\lambda, \gamma} \to \lrp{0,0}} 
        \MMSE{\Y + \gamma N}{\X + \lambda M} 
        &= \MMSE{\Y}{\X} .\quad\ 
    \end{align}
\end{cor} 

We note that the noises $M$ and $N$ in \corref{cor:additive-noises} may be dependent and even equal. 

\begin{cor}[Machine precision effect]
\label{cor:machine-precision}
    Let $X$ and $Y$ be two RVecs such that \mbox{$\normE{\Y} < \infty$}.
        Define $\lfloor x \rfloor_a \triangleq \lfloor x / a \rfloor \cdot a$ for $a > 0$. Then, 
        \begin{align}
        \label{eq:continuity:quantization}
            \lim_{\substack{\lrp{\lambda, \gamma} \to \lrp{0,0} \\ \lambda > 0, \gamma > 0}}
            \MMSE{\floor{\Y}_\gamma}{\floor{\X}_\lambda}
            &= \MMSE{\Y}{\X}.
        \end{align} 
\end{cor} 

Corollaries~\ref{cor:additive-noises} and~\ref{cor:machine-precision} 
answer in the affirmative questions~\eqref{eq:additiveNoise} and~\eqref{eq:machinePrecision} posed in Section~\ref{s:intro}.


\section{Continuity of the LMMSE}
\label{ss:linear-MMSE}

In this section, we treat the continuity in distribution of the LMMSE which is defined and characterized next~\cite[Chapter~8.4]{Gubner:Random-Processes-for-EE:Book:2006}, \cite[Chapter~9.1]{Intro-probability:MMSE-estimation:Book}, \cite[Chapter~3]{Babook:Estimation}.

\begin{defn}[LMMSE]
\label{def:LMMSE}
    The MMSE achieved by linear (affine) estimation, the LMMSE, of an RVec $\Y \in \reals^k$ from an RVec $\X \in \reals^m$
    such that 
    $\normE{X}, \normE{Y} <\infty$,
    is defined as 
    \begin{align}
        \LMMSE{\Y}{\X} \defeq \inf \normE{\Y - \lrp{A \X + b}}^2 ,
    \end{align}
    where the infimum is over all deterministic vectors $b \in \reals^k$ and deterministic matrices $A \in \reals^{k \times m}$.
\end{defn}

\begin{thm}[LMMSE and LMMSE estimator]
\label{thm:LMMSE}
    Let $X$ and $Y$ be two RVecs such that $\normE{X}, \normE{Y} < \infty$.
    Denote 
    \begin{align}
        \eta_X &\defeq \E{X}, 
        & C_X &\defeq \E{\lrp{X - \eta_X}\lrp{X - \eta_X}^T},
    \\* \eta_Y &\defeq \E{Y},
        & C_Y &\defeq \E{\lrp{Y - \eta_Y}\lrp{Y - \eta_Y}^T},
    \\* && C_{\Y,\X} &\defeq \E{\lrp{\Y - \eta_\Y}\lrp{\X - \eta_\X}^T},
    \end{align}
    and assume that $C_\X$ is invertible.\footnote{If $C_\X$ is not invertible, this means that the entries of $\X - \E{\X}$ are linearly dependent a.s. Hence, to attain the LMMSE estimator, one may remove all the linearly dependent entries and estimate from the remaining entries without loss of performance.}
    Then, the LMMSE estimate $\hY$ of $\Y$ from $\X$ is given by
    \begin{align}
        \hY = \eta_{\Y} + C_{\Y, \X} C_\X^{-1} \lrp{\X - \eta_\X}, 
    \end{align}
    i.e., 
    $A = C_{\Y, \X} C_\X^{-1}$ and $b = \eta_{\Y} - C_{\Y, \X} C_\X^{-1} \eta_\X$ in \defref{def:LMMSE}.
    The corresponding LMMSE is given as 
    \begin{subequations}
    \label{eq:thm:LMMSE}
    \begin{align}
        \LMMSE{\Y}{\X} 
        &= \normE{\Y}^2 - \normE{\hY}^2
    \label{eq:thm:LMMSE:diff-var}
     \\ &= \trace \lrc{C_\Y - C_{\Y, \X} C_\X^{-1} C_{\Y, \X}^T} .
    \label{eq:thm:LMMSE:formula}
    \end{align}
    \end{subequations}
\end{thm}
\thmref{thm:LMMSE} suggests that the LMMSE is continuous if the means, variances, and covariances are continuous. The following theorem, whose proof is available in \appref{app:LMMSE}, shows that these continuity requirements hold in case of convergence in distribution and the continuity of the individual second moments.

\begin{thm}[LMMSE continuity in distribution]
\label{thm:LMMSE:continuity}
    Let $(X,Y)$ be a pair of RVecs and let $\seq{\lrp{X_n, Y_n}}$ be a sequence of pairs of RVecs such that 
    \begin{enumerate}\addtolength\itemsep{.4\baselineskip}
    \item
    \label{itm:LMMSE:continuity:(Xn,Yn)-->(X,Y)}
        $\lrp{X_n,Y_n} \stochto{d} (X,Y)\,$;
    \item
    \label{itm:LMMSE:continuity:moment-converge}
        $\Limn \E{X_n^2} = \E{X^2} < \infty$
        and\\
        $\Limn \E{Y_n^2} = \E{Y^2} < \infty \,$.\footref{foot:entrywise-squaring}
    \item 
    \label{itm:LMMSE:continuity:Cy^-1}
        $C_\X$ is invertible.
    \end{enumerate}
    Then, 
     the LMMSE is continuous in distribution:
     \begin{align}
     \label{eq:thm:LMMSE:continuity}
         \limn \LMMSE{\Y_n}{\X_n} = \LMMSE{\Y}{\X}.
     \end{align}
\end{thm}
 
We next review examples~\ref{ex:not-m.s.} and~\ref{ex:X+Y/n} and introduce a new example to demonstrate the necessity of the requirements of \thmref{thm:LMMSE:continuity}.

\begin{example}
    Consider the setting of \exref{ex:not-m.s.}. Note that all the MMSE estimators in this example are linear, meaning that the MMSEs coincide with their corresponding LMMSEs. This demonstrates, in turn, the necessity of the convergence of the second moment of $\seq{\Y_n}$ to that of $\Y$ in requirement~\ref{itm:LMMSE:continuity:moment-converge}
    of \thmref{thm:LMMSE:continuity}.
\end{example}

\begin{example}
    Let $\Y$ be some RV with zero mean and unit variance. Set $\X = \Y$, $\Y_n = \Y$ for all $n \in \nats$, and
    \begin{align}
       \X_n = 
        \begin{cases}
            \sqrt{n}, & \mathrm{w.p.}\ \frac{1}{2n}
         \\ -\sqrt{n}, & \mathrm{w.p.}\ \frac{1}{2n}
         \\ \Y, & \mathrm{w.p.}\ 1 - \frac{1}{n} ,
        \end{cases}
    \end{align}
    where the mixture branch and its equiprobable sign are chosen independently of $\Y$.
    Clearly, requirements~\ref{itm:LMMSE:continuity:(Xn,Yn)-->(X,Y)} and~\ref{itm:LMMSE:continuity:Cy^-1} of \thmref{thm:LMMSE:continuity} hold for all $n \in \nats$. Moreover, 
    \begin{align}
        \Limn \E{\Y^2_n} = \Limn 1 = 1 = \E{\Y^2}.
    \end{align}
    
    However, 
    \begin{align}
        \limn \E{\X_n^2} = 2 > 1 = \E{\X^2}.
    \end{align}
    Hence the second part of requirement~\ref{itm:LMMSE:continuity:moment-converge} 
    of \thmref{thm:LMMSE:continuity} is violated. Indeed, using the standard formula for the LMMSE (see \thmref{thm:LMMSE}) yields
    \begin{align}
        \begin{aligned}
        \limn \LMMSE{\Y_n}{\X_n}
        &= \limn \lrp{\E{\Y_n^2} - \frac{\Esqr{\Y_n \X_n}}{\E{\X_n^2}}}\\
        &= 1 - \frac{1}{2} > 0 = \LMMSE{\Y}{\X} ,
        \end{aligned}
    \end{align}
    which demonstrates the necessity of the second part of requirement~\ref{itm:LMMSE:continuity:moment-converge} in \thmref{thm:LMMSE:continuity}.
\end{example}

\begin{example}
    Consider the setting of \exref{ex:X+Y/n}. 
    By \thmref{thm:LMMSE:continuity}, the LMMSE is continuous, as long as $\Var{\X}>0$. 
    The discrepancy between the continuity of the LMMSE and the discontinuity of the MMSE stems from the linearity constraint of the LMMSE estimator, as the perfect recovery of $\Y$ from $\X_n$ is non-linear.
    However, for $\X = 0$ and $\Y$ such that $\var{\Y} > 0$:
    \begin{align}
        \LMMSE{\Y_n}{\X_n} &= 0 &\forall n \in \nats ,
     \\ \LMMSE{\Y}{\X} &= \Var{\Y} > 0.
    \end{align}
    This, in turn, demonstrates the necessity of requirement~\ref{itm:LMMSE:continuity:Cy^-1} in \thmref{thm:LMMSE:continuity}, which is violated in this case.
\end{example}


\section{Extension: Continuous Costs}
\label{s:general-cost}

In this section, we extend the results of \secref{s:main} from a squared Euclidean distance cost function to more general costs. 

\begin{defn}[Minimal mean cost]
\label{def:opt-cost}
    Let $c: \cY \times \hat{\cY} \to \reals$ be a cost function. Then, the minimal mean cost in estimating an RVec $\Y \in \cY$ from an RVec $\X \in \cX$ is defined as 
    \begin{align}
        \ocost{\Y}{\X} \triangleq \inf \E{\cost{\Y}{\hY}} ,
    \end{align}
    where the infimum is over all RVecs $\hY$ taking values in $\hat{\cY}$ that satisfy $\Y \markov \X \markov \hY$ for which $\E{\cost{\Y}{\hY}}$ is defined.
\end{defn}

As in the case of a squared Euclidean distance cost function, we may focus on estimators that are deterministic functions of the measurements, without loss of generality, viz.
\begin{align}
\label{eq:general-cost:optimal-cost-def}
    \ocost{\Y}{\X} = \inf_{g:\ \cX \to \mathcal{\hY}} \E{c \lrp{\Y, g\lrp{\X}}} 
\end{align}
where the infimum is taken over all measurable functions \mbox{$g: \cX \to \hat{\cY}$} for which $\E{c \lrp{\Y, g\lrp{\X}}}$ is defined.

We first slightly extend a result regarding u.s.c.~\cite{Yuksel-Linder:continuity-control:SICON2012}, \cite[Theorem~8.3.3]{Yuksel-Basar:games:book:2024}
for expected continuous bounded costs, from 
$\lrp{\Y, \X_n} \stochto{d} \lrp{\Y, \X}$ to $\lrp{\Y_n, \X_n} \stochto{d} \lrp{\Y, \X}$.

\begin{thm}[u.s.c.~in distribution of bounded continuous cost]
\label{thm:gen-cost:usc:Yuksel-extended}
    Let $c: \cY \times \mathcal{\hY} \to \reals$, $\lrp{X,Y} \in \mathcal{X} \times \mathcal{Y}$ be a pair of RVecs, and $\seq{\lrp{X_n, Y_n} \in \mathcal{X} \times \mathcal{Y}}$ be a sequence of pairs of RVecs such that
    \begin{enumerate}\addtolength\itemsep{.2\baselineskip}
    \item 
        $c$ is continuous and bounded;
    \item 
        $\mathcal{\hY}$ is a convex set;
    \item 
        $\lrp{X_n, Y_n} \stochto{d} \lrp{X, Y}$.
    \end{enumerate}
    Then, the minimal mean cost is u.s.c.\ in distribution:
    \begin{align}
        \limsup_{n \to \infty} \ocost{\Y_n}{\X_n} \leq \ocost{\Y}{\X}.
    \end{align}
\end{thm}

The proof of \thmref{thm:gen-cost:usc:Yuksel-extended} is the same as that of \cite[Theorem~8.3.3]{Yuksel-Basar:games:book:2024} with $P Q_n$ replaced with $P_n Q_n$ in the last block of equations in the proof, with $P_n$ denoting the probability distribution of $X_n$.

Note that \thmref{thm:gen-cost:usc:Yuksel-extended} is limited to bounded (and continuous) cost functions meaning that it does not subsume \thmref{thm:MMSE-usc} for a squared Euclidean distance cost function, unless $\seq{\Y_n}$ is a bounded sequence, to wit, unless there exists some constant $m$, such that $\abs{\Y_n} \leq m$ a.s.\ for all $n \in \nats$.

We next consider two classes of continuous costs, which are not (necessarily) bounded: Bregman divergences in Section~\ref{ss:general-cost:Bregman-divergence} and continuous (generalized-)metric costs in Section~\ref{ss:general-cost:metric-cost}.


\subsection{Bregman Divergence Costs}
\label{ss:general-cost:Bregman-divergence}

Here we consider Bregman divergences, defined in~\defref{def:Bregman-divergence}.

\begin{defn}[Bregman divergence]
\label{def:Bregman-divergence}
    Let $\phi : \cY \to \reals$ be a strictly convex differentiable function, where $\cY$ is a convex subset of $\reals^k$ for $k \in \nats$. Then, the Bregman divergence associated with $\phi$ is defined as 
    \begin{align}
        \cost{\y_1}{\y_2} = \phi\lrp{\y_1} - \phi\lrp{\y_2} - \inner{\y_1 - \y_2}{\nabla \phi\lrp{\y_2}}.
    \end{align}
\end{defn}

Important Bregman divergences include the squared Euclidean norm generated by $\phi\lrp{x} = \norm{x}^2$ and $\cY = \reals^k$, the Kullback--Leibler divergence generated by $\phi(x) = \sum_{i=1}^k x[i] \log x[i]$ and $\cY$ being the relative interior of the $k$-simplex.\footnote{We use the convention $0 \log 0 = 0 \log \frac{0}{0} = 0 \log \frac{0}{p} = 0$ and $p \log \frac{p}{0} = \infty$ for $p > 0$, in the definition of the Kullback--Leibler divergence.}
Other important examples can be found in~\cite{Banerjee-Guo-Wang:Bregman-loss-function:estimation:TIT2005} and follow-up work.

The conditional expectation is known to be the optimal estimator for Bregman divergence cost functions~\cite{Banerjee-Guo-Wang:Bregman-loss-function:estimation:TIT2005}, as stated next.

\begin{thm}[Minimal mean Bregman divergence]
\label{thm:Bregman-divergence:optimal-estimator}
    Let $X$ and $Y$ be two random vectors where $\cY$ is a convex subset of $\reals^k$ for $k \in \nats$. Let $\phi: \cY \to \reals$ be strictly convex and differentiable, and $c$ the Bregman cost function associated with $\phi$.
    Assume that $\E{X}$ and $\E{\phi(X)}$ exist and are finite.
    Then, $\CE{\Y}{\X}$ is the unique optimal estimator of $\Y$ given $\X$ with respect to $c$, to wit,
    \begin{align}
        \ocost{\Y}{\X} = \E{\cost{\Y}{\CE{\Y}{\X}}}.
    \end{align}
\end{thm}

Theorems~\ref{thm:Markov-continuity:estimator} and~\ref{thm:Bregman-divergence:optimal-estimator} suggest the following Markovian continuity of mean Bregman cost functions.

\begin{thm}[Markovian continuity in probability of minimal mean Bregman divergence]
\label{thm:Bregman-cost:Markov-continuity}
    Let $(X,Y) \in \mathcal{X} \times \mathcal{Y}$ be a pair of RVecs, and let $\seq{\lrp{X_n, Y_n}}$ be a sequence of pairs of RVecs over $\mathcal{X} \times \mathcal{Y}$, where 
    $\cY$ is a convex subset of $\reals^k$ and $\cX \subseteq \reals^m$.
    Let $\phi: \cY \to \reals$ be strictly convex and differentiable, and $c$ the Bregman divergence cost associated with $\phi$. Assume that 
    \begin{enumerate}\addtolength\itemsep{.4\baselineskip}
    \item
        $\lrp{X_n,Y_n} \stochto{p} (X,Y)$;
    \item 
    \label{itm:Bregman:Markov-continuity:moment-convergence}
        $\Limn \E{\abs{\Y_n}} = \E{\abs{\Y}}$ and $\normLone{\Y} < \infty$;
    \item 
    \label{itm:Bregman:Markov-continuity:phi-moment-convergence}
        $\Limn \E{\abs{\phi\lrp{\Y_n}}} = \E{\abs{\phi \lrp{\Y}}} < \infty$;
    \item 
        $\Y \markov \X \markov \X_n$ for all $n \in \nats$.
    \end{enumerate}
     Then, the minimal mean Bregman divergence is Markov continuous in probability:
     \begin{align}
     \label{eq:Bregman:Markov-continuity}
         \limn \ocost{\Y_n}{\X_n} = \ocost{\Y}{\X}.
     \end{align}
\end{thm}

\begin{remark}
    The absolute values in requirements~\ref{itm:Bregman:Markov-continuity:moment-convergence} and~\ref{itm:Bregman:Markov-continuity:phi-moment-convergence} of \thmref{thm:Bregman-cost:Markov-continuity} may be dropped in many cases of interest; see \remref{rem:drop-abs} for details.
\end{remark}

To prove \thmref{thm:Bregman-cost:Markov-continuity} we first present two auxiliary results. 

\begin{lem}
\label{lem:Bregman:E[phi(hX_n)]-->E[phi(hX)]}
    Under the conditions of \thmref{thm:Bregman-cost:Markov-continuity}, 
    $\Limn \E{\phi \lrp{\CE{\Y_n}{\X_n} } } = \E{\phi \lrp{\CE{\Y}{\X}}}$.
\end{lem}

The proof of \lemref{lem:Bregman:E[phi(hX_n)]-->E[phi(hX)]} is available in \appref{app:gen-cost:usc:metric-cost}.

The next result can be found in the proof of Theorem~1 in \cite{Banerjee-Guo-Wang:Bregman-loss-function:estimation:TIT2005}, and is easily proved by applying the law of total expectation on $\E{\cost{\Y}{\CE{\Y}{\X}}}$.
\begin{lem}
\label{lem:Bregman:optimal-cost-expression}
    Let $(X,Y) \in \mathcal{X} \times \mathcal{Y}$ be a pair of RVecs, where 
    $\cY$ is a convex subset of $\reals^k$ and $\cX \subseteq \reals^m$, $\norm{\Y}_{L^1} < \infty$, and $\E{\abs{\phi(\Y)}} < \infty$.
    Let $\phi: \cY \to \reals$ be strictly convex and differentiable, and~$c$ the Bregman divergence cost associated with~$\phi$.
    Then, 
    \begin{align}
    \label{eq:Bregman:optimal-cost-expression}
        \ocost{\Y}{\X} = \E{\phi\lrp{\Y}} - \E{\phi\lrp{\CE{\Y}{\X}}}.
    \end{align}
\end{lem}

We are now ready to prove \thmref{thm:Bregman-cost:Markov-continuity}.

\begin{IEEEproof}[Proof of~\thmref{thm:Bregman-cost:Markov-continuity}]
Denote $\hY = \CE{\Y}{\X}$ and $\hY_n = \CE{\Y_n}{\X_n}$ for $n \in \nats$.
Using lemmata~\ref{lem:Bregman:E[phi(hX_n)]-->E[phi(hX)]} and~\ref{lem:Bregman:optimal-cost-expression}, we attain the desired result:
    \begin{align}
        \limn \ocost{\Y_n}{\X_n} 
        &= \limn \lrc{ \E{\phi\lrp{\Y_n}} - \E{\phi\lrp{\hY_n}} }
     \\ &= \limn \E{\phi\lrp{\Y_n}} - \limn \E{\phi\lrp{\hY_n}} 
     \\ &= \E{\phi\lrp{\Y}} - \E{\phi\lrp{\hY}} 
     \\ &= \ocost{\Y}{\X}, 
    \end{align}
    where 
    the limits exist by \lemref{lem:Bregman:E[phi(hX_n)]-->E[phi(hX)]} and requirement 3 together with the continuous mapping theorem (\thmref{thm:continuous-mapping-theorem}) and Vitali's convergence theorem (\thmref{thm:Vitali:prob+2ndMom-->m.s.}).
\end{IEEEproof}

For Kullback--Leibler divergence, the same Markov continuity conclusion holds on the closed simplex. Direct calculation shows that the conditional mean is optimal and that~\eqref{eq:Bregman:optimal-cost-expression} holds for $\phi(x) = \sum_i x[i]\log x[i]$. Since $\phi$ is bounded and continuous on the closed simplex, \thmref{thm:Markov-continuity:estimator} with $r=1$ and Vitali's convergence theorem (\thmref{thm:Vitali:prob+2ndMom-->m.s.}) establish the conclusion.

\begin{remark}
    Bregman divergence is defined with a \textit{strictly} convex $\phi$. 
    However, all the aforestated results continue to hold for a (weakly) convex $\phi$. The only difference lies in the conditional mean $\CE{\Y}{\X}$ possibly not being the only optimal estimator.
\end{remark}


\subsection{Continuous Metric and Generalized-Metric Costs}
\label{ss:general-cost:metric-cost}

We next derive a similar result for continuous metric cost functions which are not necessarily bounded. 
This family of cost functions includes the Minkowski and Chebyshev  distances---$\cost{\x_1}{\x_2} = \norm{\x_1-\x_2}_r$ for $r \in [1, \infty]$, as well as $\cost{\x_1}{\x_2} = \norm{\x_1 - \x_2}_r^r$ for $r \in (0,1)$.
For consistency with the rest of the paper, we focus on the case of finite-dimensional Euclidean spaces: $\cY, \hat{\cY} \subseteq \reals^k$ and $\cX \subseteq \reals^m$.

\begin{defn}[continuous metric cost]
\label{def:metric-cost}
    A continuous metric cost function $c : \cY \times \cY \to \reals$ over a closed subset $\cY \subseteq \reals^{k}$
    is a cost $c$ satisfying, for all $\y_1, \y_2, \y_3 \in \cY$:
    \begin{itemize}\addtolength\itemsep{.5\baselineskip}
    \item 
        \textit{Continuity:} The function $c$ is jointly continuous on $\cY \times \cY$ (with respect to the Euclidean topology):
        $$\Lim_{\lrp{\ty_1, \ty_2} \to \lrp{\y_1, \y_2}} \cost{\ty_1}{\ty_2} = \cost{\y_1}{\y_2}.$$
    \item 
        \textit{Positivity:}
        $\cost{\y_1}{\y_2} \geq 0$ with equality if and only if (iff) $\y_1 = \y_2$.
    \item 
        \textit{Symmetry:}
        $\cost{\y_1}{\y_2} = \cost{\y_2}{\y_1}$.
    \item 
        \textit{Triangle inequality:}
        $\cost{\y_1}{\y_3} \leq \cost{\y_1}{\y_2} + \cost{\y_2}{\y_3}$.
    \end{itemize}
\end{defn}

\begin{remark}
\label{rem:relaxed-gen-costs}
    In all the results to follow pertaining to continuous metric costs, the positivity property of \defref{def:metric-cost} can be relaxed to 
    $\cost{\y_1}{\y_2} \geq 0$ with equality if (but not necessarily only if) $\y_1 = \y_2$. That is, all the results extend to continuous \textit{pseudo}metric costs. 
\end{remark}

The following theorem is proved in \appref{app:gen-cost:usc:metric-cost}.
\begin{thm}[u.s.c.~in distribution of minimal mean continuous metric cost]
\label{thm:gen-cost:usc:metric-cost}
    Let $(X,Y) \in \mathcal{X} \times \mathcal{Y}$ be a pair of RVecs, and let $\seq{\lrp{X_n, Y_n}}$ be a sequence of pairs of RVecs over $\mathcal{X} \times \mathcal{Y}$ such that 
    \begin{align}
        \lrp{X_n,Y_n} \stochto{d} (X,Y) , 
    \end{align}
    where $\cY$ is a closed convex subset of $\reals^k$ and $\cX \subseteq \reals^m$.
    Let $c: \cY \times \cY \to \reals$ be a continuous metric cost function (as in \defref{def:metric-cost}), such that there exists $\oy \in \cY$ for which 
    \begin{itemize}\addtolength\itemsep{.5\baselineskip}
    \item 
        $\E{\cost{\Y_n}{\oy}}, \E{\cost{\Y}{\oy}} < \infty$ for all $n \in \nats$;
    \item 
        $\Limn \E{\cost{\Y_n}{\oy}} = \E{\cost{\Y}{\oy}}$.
    \end{itemize}
    Then, the minimal mean cost is u.s.c.\ in distribution:
    \begin{align}
    \label{eq:gen-cost:usc:metric-cost}
        \limsup_{n \to \infty} \ocost{\Y_n}{\X_n} \leq \ocost{\Y}{\X}.
    \end{align}
\end{thm}

\begin{remark}
    \thmref{thm:gen-cost:usc:metric-cost} holds for the Manhattan distance, $\cost{\x_1}{\x_2} = \norm{\x_1-\x_2}_1$, although the median---the best estimator under this cost---is not unique in general. In this context, recall also \remref{rem:relaxed-gen-costs}.
\end{remark}

Since the squared Euclidean distance $c(\y_1, \y_2) = \norm{\y_1 - \y_2}^2$ does not satisfy the triangle inequality, the continuity results established in \secref{s:main} do not fit the framework of \thmref{thm:gen-cost:usc:metric-cost}. We next slightly extend the result of \thmref{thm:gen-cost:usc:metric-cost} to support MMSE by allowing the application of a continuous strictly increasing function $f$ to expected costs such that the generalized triangle inequality below holds;
the proof of this result can be found in 
\appref{app:gen-cost:usc:metric-cost}.
\begin{cor}[u.s.c.\ in distribution of minimal mean continuous generalized-metric cost]
\label{cor:gen-cost:usc:metric-cost-after-func}
    Let $(X,Y) \in \mathcal{X} \times \mathcal{Y}$ be a pair of RVecs, and let $\seq{\lrp{X_n, Y_n}}$ be a sequence of pairs of RVecs over $\mathcal{X} \times \mathcal{Y}$, such that 
    \begin{align}
        \lrp{X_n,Y_n} \stochto{d} (X,Y),
    \end{align}
    where $\cY$ is a closed convex subset of $\reals^k$ and $\cX \subseteq \reals^m$.
    Let $f: \reals_{\geq 0} \to \reals_{\geq 0}$ be a continuous strictly increasing function
    that satisfies $f(0) = 0$. 
    Let $c: \cY \times \cY \to \reals$ and $\oy \in \cY$ such that 
    \begin{itemize}\addtolength\itemsep{.5\baselineskip}
    \item 
        $c$ satisfies the continuity, positivity and symmetry properties of \defref{def:metric-cost};
    \item 
        Generalized triangle inequality: for every jointly distributed triplet of RVecs $A_1, A_2, A_3$ taking values in $\cY$ the inequality
        \begin{align}
        \label{eq:generalized-triangle-inequality}
            f\lrp{ \E{\cost{A_1}{A_3}} } \leq f\lrp{ \E{\cost{A_1}{A_2}} } + f\lrp{ \E{\cost{A_2}{A_3}} }
        \nonumber
        \end{align}
        holds whenever the expectations are finite;
    \item 
        $\E{\cost{\Y_n}{\oy}}, \E{\cost{\Y}{\oy}} < \infty$ for all $n \in \nats$;
    \item 
        $\Limn \E{\cost{\Y_n}{\oy}} = \E{\cost{\Y}{\oy}}$.
    \end{itemize}
    Then, the minimal mean cost is u.s.c.\ in distribution:
    \begin{align}
        \limsup_{n \to \infty} \ocost{\Y_n}{\X_n} \leq \ocost{\Y}{\X}.
    \end{align}
\end{cor}

Taking $c$ to be the squared Euclidean distance and $f$ to be the square root specializes to the result of \thmref{thm:MMSE-usc} for the MMSE. Similarly, cost functions of the form $\cost{\y_1}{\y_2} = \norm{\y_2 - \y_1}_r^r$ for $r \in [1, \infty)$ fall within the framework of \corref{cor:gen-cost:usc:metric-cost-after-func}, as well as the Jensen--Shannon divergence~\eqref{eq:Jensen--Shannon-Divergence:def} over standard simplices.

The following corollary generalizes \thmref{thm:garbled:weak-continuity}.

\begin{cor}[Markovian continuity in distribution of minimal mean continuous metric cost with common parameter]
\label{cor:garbled:weak-continuity:general-cost}
    Let $(X,Y) \in \mathcal{X} \times \mathcal{Y}$ be a pair of RVecs over $\mathcal{X} \times \mathcal{Y}$, and let $\seq{\X_n}$ be a sequence of RVecs over $\cX$ such that 
    \begin{align}
        \lrp{\Y, \X_n} \stochto{d} (\Y, \X) , 
    \end{align}
    where $\cY$ is a closed convex subset of $\reals^k$ and $\cX \subseteq \reals^m$.
    Let $c: \cY \times \cY \to \reals$ be a continuous metric cost function (as in \defref{def:metric-cost}), such that there exists $\oy \in \cY$ for which:
    \vspace{.5\baselineskip}
    \begin{itemize}\addtolength\itemsep{.5\baselineskip}
    \item 
        $\E{\cost{\Y}{\oy}} < \infty$;
    \item 
        $\Y \dmarkov \X \dmarkov \X_n$ for all $n \in \nats$.
    \end{itemize}
    \vspace{.5\baselineskip}
    Then, the minimal mean cost is continuous in distribution:
    \begin{align}
    \label{eq:cor:garbled:weak-continuity:general-cost:continuity}
        \Limn \ocost{\Y}{\X_n} = \ocost{\Y}{\X}.
    \end{align}
\end{cor}

The u.s.c.\ in distribution of the minimal mean cost in \corref{cor:garbled:weak-continuity:general-cost} follows from \corref{cor:gen-cost:usc:metric-cost-after-func} (as in the proof of \thmref{thm:garbled:weak-continuity}), whereas the l.s.c.\ in distribution part follows immediately from Blackwell's informativeness theorem.

The minimal mean cost is continuous in distribution~\eqref{eq:cor:garbled:weak-continuity:general-cost:continuity} when the triangle inequality in \defref{def:metric-cost} is replaced by the generalized triangle inequality in \corref{cor:gen-cost:usc:metric-cost-after-func}.

\begin{thm}[Markovian continuity in probability of minimal mean continuous metric cost]
\label{thm:Markov-continuity:general-cost}
    Let $(X,Y) \in \mathcal{X} \times \mathcal{Y}$ be a pair of RVecs and let $\seq{\lrp{X_n, Y_n}}$ be a sequence of RVecs over $\mathcal{X} \times \mathcal{Y}$ such that 
    \begin{align}
        \lrp{\Y_n, \X_n} \stochto{p} (\Y, \X) , 
    \end{align}
    where $\cY$ is a closed convex subset of $\reals^k$ and $\cX \subseteq \reals^m$.
    Let $c: \cY \times \cY \to \reals$ be a continuous metric cost function (as in \defref{def:metric-cost}), such that there exists $\oy \in \cY$ for which:
    \vspace{.5\baselineskip}
    \begin{itemize}\addtolength\itemsep{.5\baselineskip}
    \item 
        $\E{\cost{\Y_n}{\oy}}, \E{\cost{\Y}{\oy}} < \infty$ for all $n \in \nats$;
    \item 
        $\Limn \E{\cost{\Y_n}{\oy}} = \E{\cost{\Y}{\oy}}$.
    \item 
        $\Y \markov \X \markov \X_n$ for all $n \in \nats$.
    \end{itemize}
    \vspace{.5\baselineskip}
    Then, the minimal mean cost is continuous in probability:
    \begin{align}
    \label{eq:garbled:weak-continuity:general-cost}
        \Limn \ocost{\Y_n}{\X_n} = \ocost{\Y}{\X}.
    \end{align}
\end{thm}

\begin{IEEEproof}
    By \thmref{thm:gen-cost:usc:metric-cost}, u.s.c.\ holds~\eqref{eq:gen-cost:usc:metric-cost}.
    We are left with proving l.s.c.
    Let $\eps > 0$. 
    By the theorem assumption, the continuous mapping theorem and Vitali's convergence theorem, $\Limn \E{c\lrp{\Y_n,\Y}} = 0$ (see \lemref{lem:app:gen-cost:usc:metric-cost:c(Xn,X)-->0} in \appref{app:gen-cost:usc:metric-cost} for a formal proof). 
    To wit, there exists $n_0 \in \nats$ such that
    \begin{align}
    \label{eq:garbled:weak-continuity:general-cost:proof:convergence:E[c(Yn,Y)]-->0}
        \E{c\lrp{\Y_n,\Y}} < \eps 
    \end{align}
    for all $n \geq n_0$. 
    Consequently, for all $n \geq n_0$:
    \begin{subequations}
    \label{eq:garbled:weak-continuity:general-cost:proof}
    \begin{align}
        \ocost{\Y_n}{\X_n} 
        &\triangleq \inf_g \E{\cost{\Y_n}{g \lrp{\X_n}}}
    \label{eq:garbled:weak-continuity:general-cost:proof:def-opt-cost1}
     \\ &\geq \inf_g \E{\cost{\Y}{g \lrp{\X_n}}} - \E{\cost{\Y_n}{\Y}} \ \ 
    \label{eq:garbled:weak-continuity:general-cost:proof:triangle}
     \\ &> \inf_g \E{\cost{\Y}{g \lrp{\X_n}}} - \eps
    \label{eq:garbled:weak-continuity:general-cost:proof:convergence1}
    \\ &\geq \inf_g \E{\cost{\Y}{g \lrp{\X}}} - \eps
    \label{eq:garbled:weak-continuity:general-cost:proof:blackwell}
     \\ &= \ocost{\Y}{\X} - \eps ,
    \label{eq:garbled:weak-continuity:general-cost:proof:def-opt-cost2}
    \end{align}
    \end{subequations}
    where 
    \eqref{eq:garbled:weak-continuity:general-cost:proof:def-opt-cost1} and \eqref{eq:garbled:weak-continuity:general-cost:proof:def-opt-cost2} are by \defref{def:opt-cost};
    \eqref{eq:garbled:weak-continuity:general-cost:proof:triangle} follows from the triangle inequality; 
    \eqref{eq:garbled:weak-continuity:general-cost:proof:convergence1} follows from \eqref{eq:garbled:weak-continuity:general-cost:proof:convergence:E[c(Yn,Y)]-->0}; 
    \eqref{eq:garbled:weak-continuity:general-cost:proof:blackwell} follows from
    the Markov relation $\Y \markov \X \markov \X_n$, and by Blackwell's informativeness theorem.
    Since $\eps > 0$ is arbitrary, \eqref{eq:garbled:weak-continuity:general-cost:proof} guarantees the desired Markovian l.s.c.\ in probability of the minimal mean cost.
\end{IEEEproof}

Again, the minimal mean cost remains continuous in probability~\eqref{eq:garbled:weak-continuity:general-cost} when the triangle inequality in \defref{def:metric-cost} is replaced by the generalized triangle inequality in \corref{cor:gen-cost:usc:metric-cost-after-func}. Indeed, this inequality and Blackwell's theorem give
\begin{align}
    f\lrp{\ocost{\Y}{\X}} \leq f \lrp{\ocost{\Y_n}{\X_n}} + f \lrp{\E{\cost{\Y_n}{\Y}}}, 
\end{align}
and $\Lim_{n \to \infty} \E{\cost{\Y_n}{\Y}} = 0$ as in the proof of \corref{cor:gen-cost:usc:metric-cost-after-func}. Since $f$ is increasing and continuous with $f(0) = 0$, the preceding inequality and \corref{cor:gen-cost:usc:metric-cost-after-func} establish~\eqref{eq:garbled:weak-continuity:general-cost}.


\section{Discussion and Future Work}
\label{s:summary}

This work focused on bridging the gap between the practitioners' perception that MMSE is robust and theoretical results showing discontinuity in general. 

By introducing a Markov restriction~\eqref{eq:Markov:intro} 
between the nominal parameter, the nominal measurement, and the converging measurement, we proved that MMSE is in fact continuous assuming continuity of the  second moment. Such a restriction may be of interest beyond MMSE estimation, e.g., in other inference problems.

We further established results on the upper semicontinuity in distribution and 
continuity for MMSE in estimating a finite-power parameter from a converging sequence of channels under an \textit{individual} statistical degradedness assumption of each converging channel with respect to the limit channel.

We also proved that the MMSE under linear estimation is continuous in distribution assuming again continuity of the second moments, provided the covariance matrix of $\X$ is invertible.

Finally, we extended the results to continuous metric costs.

It would be interesting to explore under what other conditions the MMSE and other mean costs are continuous, as well as extend the results to a dynamical setting~\cite{Kara-Yuksel:robustness-stochastic-control:SICON2020,Kara-Yuksel:partially-observed-stochastic--control:CDC2024,Demirci-Kara-Devran-Yuksel:POMDPs-contractions:SICON2024}.

The MMSE can be used as a building block for constructing measures of statistical dependence. For geometry-oriented measures of dependence, stochastic continuity was put forward by M\'ori and Sz\'ekely~\cite{Mori-Szekely:4axioms-dependence:Metrika2019} as an axiom that such a dependence measure needs to satisfy. Since MMSE is not continuous in probability, $L^p$ (for $p \geq 1$) or a.s., it was claimed that MMSE-based dependence measures are ill-fitted as measures of dependence. Our work proves that, in fact, if the stochastic continuity axiom of M\'ori and Sz\'ekely~\cite{Mori-Szekely:4axioms-dependence:Metrika2019} is modified to require Markov continuity, then MMSE-based dependence measures~\cite{Domanovitz-Erez:dependence:ISIT2019,Nitzan_MSc} may be good measures of dependence.

The present work concentrated on the Bayesian setting where the estimated parameter(s) $\Y$ and $\seq{\Y_n}$ are RVecs. 
That said, the Markov condition $\Y \markov \X \markov \X_n$ was imposed with respect to the limit parameter $\Y$ only. Developing an analogous setting under the frequentist framework and establishing continuity results with respect to the class of probability distributions is an interesting research direction to explore.


\appendices

\section{Background on Stochastic Convergences and Stochastic Degradedness / Garbling} 
\label{app:basics}

\subsection{Stochastic Convergences}
\label{app:basics:convergence}

We first present four standard definitions of stochastic convergence \cite[Chapter~5]{Gut:Probability:Book2005}, \cite[Chapter~2]{van-der-Vaart:Asymptotic-Statistics:Book1998}.

\begin{defn}[stochastic convergences]
\label{def:convergences}
    Let $\seq{X_n}$ be a sequence of RVecs in $\reals^k$, 
    and let $X$ be an RVec, defined on the same probability space $\lrp{\Omega, \cF, \mathrm{P}}$.
    Then,
    \begin{enumerate}
    \item 
        \textit{Convergence in distribution.} 
        $\seq{X_n}$ converges in distribution to $X$
        if\footnote{Recall that `$\leq$' between vectors is interpreted componentwise.}
        \begin{align}        
            \Limn \PR{X_n \leq x} = \PR{X \leq x}        
        \end{align}
        for every $x \in \reals^k$ at which the cumulative distribution function of $X$, $\PR{X \leq x}$, is continuous at $x$.
        We denote this convergence by $X_n \stochto{d} X$.
    \item 
        \textit{Convergence in probability.} 
        $\seq{X_n}$ converges in probability to $X$ 
        if, for all $\eps > 0$,\footnote{\label{foot:metric}Other metrics between $X_n$ and $X$ that give the same convergent sequences on $\reals^k$ may be used as well.}
        \begin{align}        
            \Limn \PR{\norm{X_n - X} > \eps} = 0.        
        \end{align}
        We denote this convergence by $X_n \stochto{p} X$.
    \item 
        \textit{Almost-sure convergence.} 
        $\seq{X_n}$ converges almost surely (a.s.) to $X$ 
        if
        \begin{align}        
            \PR{\Limn X_n = X} = 1        
        \end{align}
        or, equivalently, if\footref{foot:metric}
        \begin{align}
            \PR{\Limn \norm{X_n - X} = 0} = 1 .
        \end{align}
        We denote this convergence by $X_n \stochto{a.s.} X$.
    \item
    \label{itm:convergences:m.s.}
        \textit{Convergence in mean.} 
        $\seq{X_n}$ converges in $r$-th mean to $X$ for $r \in [1, \infty)$
        if $\normEr{X} < \infty$, $\normEr{X_n} < \infty$ for all $n \in \nats$, and 
        \begin{align}        
            \Limn \normEr{X_n - X} = 0.
        \end{align}
        We denote this convergence by $X_n \stochto{L^r} X$. For the special case of $r=1$ and $r = 2$, we refer to this convergence as convergence in mean and in mean square (m.s.). We denote $X_n \stochto{L^2} X$ by $X_n \stochto{m.s.} X$.
    \end{enumerate}
\end{defn}

The following is an alternative definition of convergence in distribution, also known as \textit{convergence in law} or \textit{weak convergence}~\cite[Chapter~5, Definition~1.5]{Gut:Probability:Book2005}.
\begin{defn}[weak convergence]
\label{def:convergence:weak}
    Let $\seq{X_n}$ be a sequence of RVecs in $\reals^k$, 
    and let $X$ be an RVec in $\reals^k$. 
    Then, $\seq{X_n}$ converges in distribution to $X$ ($X_n \stochto{d} X$) if 
    \begin{align}
        \Limn \E{f\lrp{X_n}} = \E{f\lrp{X}}
    \end{align}
    for all bounded and continuous functions $f: \reals^k \to \reals$.
\end{defn}

The equivalence of the two definitions for convergence in distribution 
is often presented as part of the \textit{Portmanteau lemma}~\cite[Chapter~2]{van-der-Vaart:Asymptotic-Statistics:Book1998}, \cite[Chapters~2 and~3]{Billingsley:convergence-probability-measures:Book2013}.

\begin{remark}
    While convergences in probability, in m.s., and a.s.\ require the RVecs in $\seq{X_n}$ and $X$ to be defined on the same probability space, this is not necessary for convergence in distribution.
\end{remark}

\begin{lem}
\label{lem:convergence:subseq<=>seq}
    Let $k \in \nats$ and let $r \in [1, \infty)$. 
    Let $\seq{X_n}$ be a sequence of RVecs in $\reals^k$, 
    and let $X$ be a RVec in $\reals^k$, defined on the same probability space $\lrp{\Omega, \cF, P}$.
    Then,
    \begin{align*}
     \mathrm{a)} \ X_n &\stochto{d} X &\Rightarrow&& X_n[i] &\stochto{d} X[i] \ \ \forall i \in \firstnats{k};
     \\ \mathrm{b)} \ X_n &\stochto{p} X &\Leftrightarrow&& X_n[i] &\stochto{p} X[i] \ \ \forall i \in \firstnats{k};
     \\ \mathrm{c)} \ X_n &\stochto{a.s.} X &\Leftrightarrow&& X_n[i] &\stochto{a.s.} X[i] \ \ \forall i \in \firstnats{k};
     \\ \mathrm{d)} \ X_n &\stochto{L^r} X &\Leftrightarrow&& X_n[i] &\stochto{L^r} X[i] \ \ \forall i \in \firstnats{k}.     
    \end{align*}
\end{lem}

\begin{IEEEproof}
    a) Follows immediately from the definition of convergence in distribution.
    
    \noindent
    b) See \cite[Theorem~2.7]{van-der-Vaart:Asymptotic-Statistics:Book1998}.
    \\
    c) Define the events $A_\ell = \lrc{\Limn X_n[\ell] = X[\ell]}$ for all $\ell \in \lrc{1, 2, \ldots, k}$ and $A = \lrc{\Limn X_n = X}$. Clearly, $A = \bigcap\limits_{\ell=1}^k A_\ell$.
    
    Assume first that $X_n \stochto{a.s.} X$ and set some $i \in \lrc{1, 2, \ldots k}$. 
    Then, 
    \begin{align}
        1 &= \PR{A} 
        = \PR{\bigcap_{\ell=1}^k A_\ell}
        \leq \PR{A_i}
        \leq 1.
    \nonumber
    \end{align}
    Thus, by the squeeze theorem, $X_n[i] \stochto{a.s.} X[i]$ for all $i \in \lrc{1, 2,\ldots, k}$.

    Now assume $X_n[i] \stochto{a.s.} X[i] \ \ \forall i \in \lrc{1, 2, \ldots, k}$. Then, 
    \begin{align}
        1 &\geq \PR{A} 
        = \PR{\bigcap_{\ell=1}^k A_\ell}
        = 1 - \PR{\bigcup_{\ell=1}^k A_\ell^c}
        \geq 1 - \sum_{\ell=1}^k \PR{A^c_\ell}
        \geq 1.
    \nonumber
    \end{align}
    Thus, by the squeeze theorem, $X_n \stochto{a.s.} X$.

\noindent
    d) The result immediately follows by noting that 
    \begin{align}
        \normEr{X - X_n}^r = \sum_{i=1}^k \normEr{X[i] - X_n[i]}^r.
    \end{align}
\end{IEEEproof}

Continuous mappings preserve stochastic convergence, as is stated in the following theorem~\cite[Theorem~2.3]{van-der-Vaart:Asymptotic-Statistics:Book1998}.
\begin{thm}[Continuous mapping theorem]
\label{thm:continuous-mapping-theorem}
    Let $k,m \in \nats$. 
    Let $\seq{X_n}$ be a sequence of RVecs in $\reals^k$, 
    and let $X$ be an RVec in $\reals^k$, defined on the same probability space $\lrp{\Omega, \cF, P}$.
    Let $\phi: \reals^k \to \reals^m$ be a continuous map. 
    Then, the following relations hold:
\vspace{.2\baselineskip}
    \begin{enumerate}\addtolength\itemsep{.3\baselineskip}   
    \item 
        $X_n \stochto{d} X \ \Rightarrow\  \phi\lrp{X_n} \stochto{d} \phi\lrp{X}$; 
    \item 
        $X_n \stochto{p} X \ \Rightarrow\  \phi\lrp{X_n} \stochto{p} \phi\lrp{X}$; 
    \item 
        $X_n \stochto{a.s.} X \ \Rightarrow\  \phi\lrp{X_n} \stochto{a.s.} \phi\lrp{X}$; 
    \end{enumerate}
\end{thm}

The results in the following lemma and theorem are well known; see \cite[Chapter~5, Theorems~3.1 and~5.4]{Gut:Probability:Book2005}.
\begin{lem}
\label{lem:convergence:a.s,m.s=>p=>d}    
    Let $k \in \nats$ and $r \geq 1$. 
    Let $\seq{X_n}$ be a sequence of RVecs in $\reals^k$, 
    and let $X$ be an RVec in $\reals^k$, defined on the same probability space $\lrp{\Omega, \cF, P}$.
    Then, the following relations hold:
\vspace{.2\baselineskip}
    \begin{enumerate}\addtolength\itemsep{.3\baselineskip}   
    \item 
        $X_n \stochto{a.s.} X \ \Rightarrow\  X_n \stochto{p} X$; 
    \item 
        $X_n \stochto{L^r} X \ \Rightarrow\  X_n \stochto{p} X$; 
    \item
    \label{itm:lem:convergence:p=>d}
        $X_n \stochto{p} X \ 
        \Rightarrow\  X_n \stochto{d} X$.
    \end{enumerate}
\end{lem}

\lemref{lem:convergence:a.s,m.s=>p=>d} states that convergence in $r$-th mean guarantees convergence in probability; the opposite direction does not hold in general. However, under uniform integrability of $\seq{\abs{X_n}^r}$, defined next, the opposite direction holds as well.

\begin{defn}[uniform integrability {\cite[Chapter~5, Section~4]{Gut:Probability:Book2005}}]
\label{def:uniform-integrability}
    A sequence $\lrcm{\Y_n \in \reals}{n \in \nats}$ of RVs is said to be \textit{uniformly integrable (u.i.)} if
    for every $\eps > 0$, there exists a constant $a > 0$ 
    such that 
    \begin{align}
        \E{\indicator{\abs{\Y_n} > a} \cdot \abs{\Y_n}} &\leq \eps & \forall n \in \nats, 
    \end{align}
    where $\indicator{\cdot}$ is the indicator function.
    A sequence of RVecs $\lrcm{\Y_n \in \reals^k}{n \in \nats}$ for $k \in \nats$ is said to be u.i.\ if $\lrcm{\Y_n[i]}{n \in \nats}$ is u.i.\ for all $i \in \firstnats{k}$.
\end{defn}

The following theorem aggregates standard results regarding u.i.
\begin{thm}
\label{thm:u.i.:properties}
    Let $k \in \nats$, and let $\seq{X_n}$, $\seq{Y_n}$, and $\seq{Z_n}$ be three sequences of RVecs in $\reals^k$, such that $\seq{X_n}$ and $\seq{Z_n}$ are u.i. Then,
    \begin{enumerate}
    \item 
        $\seq{\abs{X_n}}$ is u.i.
    \item 
        $\seq{\abs{A X_n + b}}$ is u.i.\ for any $\ell \in \nats$ and deterministic matrix $A$ of dimensions $\ell \times k$ and vector $b$ of dimensions $\ell \times 1$.
    \item 
        $\seq{X_n + Z_n}$ is u.i.
    \item 
        If $\abs{Y_n} \leq Z_n$ a.s.\ for all $n \in \nats$, then $\seq{Y_n}$ is u.i.
    \item 
        If $\abs{Y_n} \leq m$ a.s.\ for all $n \in \nats$ for some constant $m \in \reals$, then $\seq{Y_n}$ is u.i.
    \item 
        If $X_n \leq Y_n \leq Z_n$ a.s.\ for all $n \in \nats$, then $\seq{Y_n}$ is u.i.
    \end{enumerate}
\end{thm}
\begin{IEEEproof}
    Property~1 holds by the definition of u.i.
    Property~2 can be easily verified by definition. 
    Properties~3--5 are standard; see~\cite[Chapter 5, Section~4]{Gut:Probability:Book2005}.
    To prove property~6, note that 
    \begin{align}
        \abs{Y_n} \leq \abs{X_n} + \abs{Z_n}. 
    \end{align}
    By properties~1 and~3, $\seq{\abs{X_n} + \abs{Z_n}}$ is u.i. Hence, by Property~4, $\seq{Y_n}$ is u.i.
\end{IEEEproof}

\begin{thm}[see {\cite[Section~3.4]{Billingsley:convergence-probability-measures:Book2013}}]
\label{thm:conv-dist:2ndMom<==>u.i.}
    Let $\seq{\Y_n}$ be a sequence of RVecs, and let $\Y$ be an RVec such that $\Y_n \stochto{d} \Y$. 
    Then, the following statements are equivalent for $r \in [1, \infty)$:
\vspace{.1\baselineskip}
    \begin{enumerate}\addtolength\itemsep{.3\baselineskip}
    \item 
        $\normEr{\Y} < \infty$, $\normEr{\Y_n} < \infty$ for all $n \in \nats$, and 
        \begin{align}
            \Limn \E{\abs{\Y_n}^r} = \E{\abs{\Y}^r};
        \end{align}
    \label{thm:conv-dist:2ndMom<==>u.i.:2ndMom}
    \item 
        $\seq{\abs{\Y_n}^r}$ is u.i.
    \label{thm:conv-dist:2ndMom<==>u.i.:u.i.}
    \end{enumerate}
\vspace{.1\baselineskip}
    Furthermore, if one of the above statements holds, then 
        $\Limn \E{\Y_n} = \E{\Y}$. 
\end{thm}

\begin{thm}[Vitali's convergence theorem~{\cite[Chapter~5, Section~5.2, Th. 5.4]{Gut:Probability:Book2005}}]
\label{thm:Vitali:prob+2ndMom-->m.s.}
    Let $\seq{\Y_n}$ be a sequence of RVecs, and let $\Y$ be an RVec. 
    Then, the following statements are equivalent for $r \in [1, \infty)$:
\vspace{.1\baselineskip}
    \begin{enumerate}\addtolength\itemsep{.3\baselineskip}
    \item 
        $\Y_n \stochto{p} \Y$, $\normEr{\Y} < \infty$, $\normEr{\Y_n} < \infty$ for all $n \in \nats$, and 
        $\Limn \E{\abs{\Y_n}^r} = \E{\abs{\Y}^r}$\ ;
    \item 
        $\Y_n \stochto{p} \Y$ and $\seq{\abs{\Y_n}^r}$ is u.i.
    \item 
        $\Y_n \stochto{L^r} \Y$;
    \end{enumerate}
\vspace{.1\baselineskip}
    Furthermore, if one of the above statements holds, then 
        $\Limn \E{\Y_n} = \E{\Y}$. 
\end{thm}

\begin{remark}
\label{rem:drop-abs}
    In both Theorems~\ref{thm:conv-dist:2ndMom<==>u.i.} and~\ref{thm:Vitali:prob+2ndMom-->m.s.}, Property~\ref{thm:conv-dist:2ndMom<==>u.i.:2ndMom} guarantees the weaker requirement $\Limn \E{\Y^r_n} = \E{\Y^r}$, for $r \in \nats$. The two are trivially equivalent for $r \in 2 \nats$. For $r \in 2 \nats + 1$ they are equivalent if $\seq{\Y_n}$ is bounded from below, by Fatou's lemma~\cite[Chapter~2, Theorem~5.2]{Gut:Probability:Book2005}.
    In fact, if $\seq{\Y_n}$ is non-negative, then $\Limn \E{\Y^r_n} = \E{\Y^r}$ is equivalent to Property~\ref{thm:conv-dist:2ndMom<==>u.i.:2ndMom}, also for non-integer $r \geq 1$.
    Furthermore, 
    by \defref{def:uniform-integrability}, if $\Y_n^r$ and $\Y^r$ are well-defined and real then the absolute values in Property~\ref{thm:conv-dist:2ndMom<==>u.i.:u.i.} can be dropped.
\end{remark}

The next lemma guarantees that if $\seq{\Y_n}$ is u.i., then the conditional expectations of the elements thereof form also a u.i.\ sequence. 

\begin{lem}[u.i.\ and conditional expectation~{\cite[E3.13]{Williams:Martingales:Book1991}}]
\label{lem:u.i.-of-conditional-expectation}
    Let $\seq{X_n}$ and $\seq{Y_n}$ be sequences of RVecs. If $\seq{\Y_n}$ is u.i., then 
    $\lrc{\CE{\Y_n}{\X_m}}_{n,m \in \nats}$ is also u.i.;
    in particular, $\seq{\CE{\Y_n}{\X_m}}$ is also u.i.
\end{lem}

The following result allows to lift convergence in distribution (or probability) to a.s.\ convergence.
\begin{thm}[Skorokhod's theorem; see {\cite[Chapter~1, Section~6]{Billingsley:convergence-probability-measures:Book2013}}]
\label{thm:Skorokhod}
    Let $X$ be an RVec and let $\lrcm{ X_n }{ n \in \nats}$ be a sequence of RVecs such that
    \begin{align}
        X_n \stochto{d} X .
    \end{align}
    Then, there exists an RVec $\tilde{X}$ and sequence of RVecs $\lrcm{ \tilde{X}_n }{ n \in \nats}$, 
    all defined on the same probability space, such that
    \begin{subequations}
    \noeqref{eq:Skorokhod}
    \label{eq:Skorokhod}
    \begin{align} 
        \tilde{X}_n  &\stackrel{d}= X_n & \forall n &\in \nats, 
    \label{eq:Skorokhod:CDF:Xn}
     \\ \tilde{X} &\stackrel{d}= X, 
    \label{eq:Skorokhod:CDF:X}
     \\ \tilde{X}_n  &\stochto{a.s.} \tilde{X} .
    \label{eq:Skorokhod:a.s.}
    \end{align}
    \end{subequations}    
\end{thm}

\subsection{Stochastic Degradedness / Garbling}

The following notion of \textit{stochastic degradedness} or \textit{garbling} and theorem will be used in the derivation of  continuity in distribution results in 
\secref{s:main}.
We define this notion in terms of RVecs rather than in the more common terms of conditional distributions~\cite[Definition~7.3.1]{Yuksel-Basar:games:book:2024}.

\begin{defn}
\label{def:garbled-full}
    Let $\lrp{X_1, Y_1}$ and $\lrp{X_2, Y_2}$ be two pairs of RVecs.
    We say that $\lrp{X_2, Y_2}$ is \textit{stochastically degraded} or \textit{garbled} with respect to $\lrp{X_1, Y_1}$ if
    \begin{align}
        \Y_1 \stackrel{d}= \Y_2
    \end{align}
    and there exist $\tY, \tX_1$, $\tX_2$ such that 
    \begin{subequations}
    \label{eq:def:garble-full}
    \begin{align}
        \lrp{\Y_1, \X_1} &\stackrel{d}= \lrp{\tY, \tX_1},
     \\ \lrp{\Y_2, \X_2} &\stackrel{d}= \lrp{\tY, \tX_2},
    \label{eq:def:garble-full:dist}
     \\ \tY &\markov \tX_1 \markov \tX_2.
    \label{eq:def:garble-full:Markov}
    \end{align}
    \end{subequations}
\end{defn}
This definition means that we can view these two pairs as two channels with the same input $\tY$ where the channel to the first output $\tX_1$ is more informative than that to the second output $\tX_2$. 
Since we are interested only in the marginal distributions of the pairs $\lrp{X_1, Y_1}$ and of $\lrp{X_2, Y_2}$ but not the joint distribution of the quadruple, we can specialize \defref{def:garbled-full} to the following. 

\begin{defn}
\label{def:garbled}
    Let $\Y, \X_1, \X_2$ be three RVecs. We will say that, given $\Y$, $\X_2$ is \textit{stochastically degraded} or \textit{garbled} with respect to $\X_1$ if there exists an RVec $\tX_1$ such that:
    \begin{subequations}
    \label{eq:def:garble}
    \begin{align}
        \lrp{\Y,\tX_1} &\stackrel{d}= \lrp{\Y,\X_1},
    \label{eq:def:garble:dist}
     \\ \Y &\markov \tX_1 \markov \X_2.
    \label{eq:def:garble:Markov}
    \end{align}
    \end{subequations}
    We denote this by $\Y \dmarkov \X_1 \dmarkov \X_2$.
\end{defn}

The following simple result can be viewed as a specialization of Blackwell's informativeness theorem \cite[Chapter~7.3.1]{Yuksel-Basar:games:book:2024} for MMSEs. Its proof is an adaptation of \cite[Theorem~11]{Wu-Verdu:MMSE:IT2012}.
\begin{lem}
\label{lem:Blackwell}
    Let $\Y \dmarkov \X_1 \dmarkov \X_2$ with $\normE{\Y} < \infty$.
    Then, 
    \begin{align}
    \label{eq:Blackwell:MMSE}
        \MMSE{\Y}{\X_1} \leq \MMSE{\Y}{\X_2},
    \end{align}
    with equality iff 
    $\CE{\Y}{\X_1} \stackrel{d}= \CE{\Y}{\X_2}$.
\end{lem}

\begin{IEEEproof}
    By \defref{def:garbled}, there exists an RVec $\tX_1$ such that $\lrp{\Y, \tX_1} \stackrel{d}{=} \lrp{\Y, \X_1}$ and $\Y \markov \tX_1 \markov \X_2$.
    Then, 
    \begin{subequations}
    \label{eq:proof:Blackwell}
    \begin{align}
        &\MMSE{\Y}{\X_2} - \MMSE{\Y}{\X_1}\nonumber\\
        &\quad= \MMSE{\Y}{\X_2} - \MMSE{\Y}{\tX_1}
    \label{eq:proof:Blackwell:i.d.}
     \\ &\quad = \normE{\CE{\Y}{\tX_1}}^2 - \normE{\CE{\Y}{\X_2}}^2
    \label{eq:proof:Blackwell:MMSE-expression}
     \\ &\quad = \normE{\CE{\Y}{\tX_1} - \CE{\Y}{\X_2}}^2
    \label{eq:proof:Blackwell:orthogonality}
     \\ &\quad \geq 0 ,
    \label{eq:proof:Blackwell:>0}
    \end{align}
    \end{subequations}
    with equality in~\eqref{eq:proof:Blackwell:>0} iff $\CE{\Y}{\tX_1} = \CE{\Y}{\X_2}$ a.s., 
    where \eqref{eq:proof:Blackwell:i.d.} follows from~$\lrp{\Y, \tX_1} \stackrel{d}{=} \lrp{\Y, \X_1}$,
    \eqref{eq:proof:Blackwell:MMSE-expression} follows from \thmref{thm:MMSE},
    and 
    \eqref{eq:proof:Blackwell:orthogonality} follows from the orthogonality principle and $\Y \markov \tX_1 \markov \X_2$, since the latter implies $\CE{\Y}{\tX_1, \X_2} = \CE{\Y}{\tX_1}$ a.s. Further, since $\lrp{\Y, \tX_1} \stackrel{d}{=} \lrp{\Y, \X_1}$, equality holds only if 
    $\CE{\Y}{\X_1} \stackrel{d}= \CE{\Y}{\X_2}$. Conversely, equality in distribution makes the difference in~\eqref{eq:proof:Blackwell:MMSE-expression} zero, concluding the proof.
\end{IEEEproof}


    





\section{Proof of \thmref{thm:MMSE-usc}}
\label{app:thm:MMSE-usc}

    Denote the dimensions of $\Y$ (and hence also of $\Y_n$) by $k \in \nats$, and that of $\X$ (and hence also of $\X_n$) by $m \in \nats$.
    Define the functions 
    \begin{subequations}
    \label{eq:usc:MMSE_estimators:g-g_n}
    \noeqref{eq:usc:MMSE_estimator_gn}
    \begin{align}
        g_n\lrp{\x} &\triangleq \CE{\Y_n}{\X_n = \x},
    \label{eq:usc:MMSE_estimator_gn}
     \\ g\lrp{\x} &\triangleq \CE{\Y}{\X = \x}.
    \label{eq:usc:MMSE_estimator_g}
    \end{align}
    \end{subequations}
    Denote the $i$-th element (scalar-valued function) of the vector-valued function $g_n$ by $g_n[i]$, 
    and, similarly, the $i$-th element (scalar-valued function) of the vector-valued function $g$ by~$g[i]$.

    Set $\eps > 0$, however small.
    By Skorokhod's theorem (\thmref{thm:Skorokhod}), there exist $\tilde{X}$, $\tilde{Y}$, $\seq{\tilde{X}_n}$, and $\seq{\tilde{Y}_n}$ that satisfy:
    \begin{subequations}
    \label{eq:MMSE-usc:Skorokhod}
    \noeqref{eq:MMSE-usc:Skorokhod:a.s.}
    \begin{align} 
        \lrp{\tilde{X}_n, \tilde{Y}_n} &\stackrel{d}= \lrp{X_n, Y_n} & \forall n &\in \nats, 
    \label{eq:MMSE-usc:Skorokhod:CDF:Xn}
     \\ \lrp{\tilde{X}, \tilde{Y}} &\stackrel{d}= \lrp{X, Y}, 
    \label{eq:MMSE-usc:Skorokhod:CDF:X}
     \\ \lrp{\tilde{X}_n, \tilde{Y}_n} &\stochto{a.s.} \lrp{\tilde{X}, \tilde{Y}}.
    \label{eq:MMSE-usc:Skorokhod:a.s.}
    \end{align}
    \end{subequations}    
    Furthermore, since 
    \begin{align}
        \Limn \E{\tY_n^2} = \Limn \E{\Y_n^2} 
        = \E{\Y^2} = \E{\tY^2} < \infty,\footref{foot:entrywise-squaring}
    \end{align}
    $\tY_n \stochto{m.s.} \tY$ by \thmref{thm:Vitali:prob+2ndMom-->m.s.}. Hence, by the definition of m.s.\ convergence (part~\ref{itm:convergences:m.s.} of \defref{def:convergences}), there exists 
    $n_1 \in \nats$ such that, for all $n > n_1$, 
    \begin{align}
    \label{eq:MMSE-usc:tY_n-tY}
        \normE{\tY_n - \tY} &< \frac{\eps}{3}. 
    \end{align}
    
    Let $i \in \firstnats{k}$ and denote 
    \begin{align}
    \label{eq:L2}
        L^2_{\X} \lrp{\reals^m} \triangleq \lrc{g: \reals^m \to \reals~\mathrm{measurable}:\ \E{g^2(\X)} < \infty}.
    \end{align}
    Denote by $C_c\lrp{\reals^m} \subset L^2_{\X} \lrp{\reals^m}$ the space of compactly-supported continuous functions (and hence also bounded) on~$\reals^m$.

    Clearly $g[i] \in L^2_{\X} \lrp{\reals^m}$ for all $i \in \firstnats{k}$ since 
    \begin{align} 
        \normE{g\lrp{\X}} &\stackrel{\text{(a)}}{=} \normE{\CE{\Y}{\X}} \stackrel{\text{(b)}}{\leq} \normE{\Y} 
         \stackrel{\text{(c)}}{<} \infty, 
    \label{eq:g^2<oo:final}
    \end{align} 
    where (a) holds by the definition of $g$~\eqref{eq:usc:MMSE_estimator_g}, 
    (b) follows from the (conditional) Jensen inequality~\cite[Chapter~9.7]{Williams:Martingales:Book1991} and the convexity of the norm function, and (c) holds by the theorem's assumption.
    
    Since $C_c \lrp{\reals^m}$ is dense in $L^2_{\X} \lrp{\reals^m}$~\cite[Theorem~3.14]{Rudin:Real-Complex-Analysis:Book1987}, 
    there exists a function
    \begin{align}
    \label{eq:MMSE_estimator_bounded}
        \hg[i] \in C_c\lrp{\reals^m} ,
    \end{align} 
    such that 
    \begin{align} 
    \label{eq:Cc_dense_L2:i} 
        \normE{g[i]\lrp{\X} - \hg[i]\lrp{\X}}
        = \normE{g[i](\tX) - \hg[i](\tX)} < \frac{\eps}{3k}.\quad
    \end{align}
    Set $\hg = \lrp{\hg[1], \hg[2], \ldots, \hg[k]}$. 
    Then, by the triangle inequality, 
    \begin{align} 
    \label{eq:Cc_dense_L2} 
        \normE{g(\tX) - \hg(\tX)} < \frac{\eps}{3} . 
    \end{align}

    By \eqref{eq:MMSE-usc:Skorokhod:a.s.}, $\tX_n \stochto{a.s.} \tX$ (recall \lemref{lem:convergence:subseq<=>seq}). Therefore, by the continuous mapping theorem (\thmref{thm:continuous-mapping-theorem}), 
    \begin{align}
        \hg \left( \tX_n \right) \stochto{a.s.} \hg(\tX) .
    \end{align}    
    Since $\hg[i] \in C_c \lrp{\reals^m}$, 
    so is $\abs{\hg[i]}^r \in C_c \lrp{\reals^m}$ for any $r > 0$. Hence, $\seq{\abs{\hg\lrp{\tX_n}}^r}$ is u.i.\ (recall \thmref{thm:u.i.:properties}; alternatively, it follows from the bounded convergence theorem).
    By \thmref{thm:Vitali:prob+2ndMom-->m.s.} (and \lemref{lem:convergence:a.s,m.s=>p=>d}), for $r = 2$:
    \begin{align}
    \label{eq:hg(Yn)-->hg(Y):L2}
        \hg \left( \tX_n \right) \stochto{L^2} \hg(\tX) ; 
    \end{align}
    that is, there exists $n_2 \in \nats$ such that, for all $n > n_2$, 
    \begin{align}
    \label{eq:hg(Xn)-->hg(X)_m.s.}
        \normE{\hg \lrp{\tX} - \hg \lrp{\tX_n}} < \frac{\eps}{3}. 
    \end{align}
    
    We are now ready to prove the desired result.
    For all $n > \max \lrc{n_1, n_2}$:
    \begin{subequations}
    \label{eq:MMSE-usc}
    \begin{align}
        &\sqrt{\MMSE{\Y_n}{\X_n}} = \sqrt{\MMSE{\tY_n}{\tX_n}}
    \label{eq:MMSE-usc:Skorokhod1}
     \\ &\qquad= \normE{\tY_n - g_n\lrp{\tX_n}}
    \label{eq:MMSE-usc:MMSE-def1}
     \\ &\qquad\leq \normE{\tY_n - \hg\lrp{\tX_n}}
    \label{eq:MMSE-usc:MMSE-opt}
     \\ &\qquad\leq \normE{\hg\lrp{\tX} - \hg\lrp{\tX_n}} + \normE{g\lrp{\tX} - \hg\lrp{\tX}}
     \\ &\qquad\quad +  \normE{\tY - g\lrp{\tX}} + \normE{\tY_n - \tY}
    \label{eq:MMSE-usc:triangle-ineq}
     \\ &\qquad< \normE{\tY - g\lrp{\tX}} + \eps
    \label{eq:MMSE-usc:epsilons}
     \\ &\qquad= \sqrt{\MMSE{\tY}{\tX}} + \eps
    \label{eq:MMSE-usc:MMSE-def2}
     \\ &\qquad= \sqrt{\MMSE{\Y}{\X}} + \eps,
    \label{eq:MMSE-usc:Skorokhod2}
    \end{align}
    \end{subequations}
    where 
    \eqref{eq:MMSE-usc:Skorokhod1} follows from \eqref{eq:MMSE-usc:Skorokhod:CDF:Xn},
    \eqref{eq:MMSE-usc:MMSE-def1} and \eqref{eq:MMSE-usc:MMSE-def2} follow from \thmref{thm:MMSE}, 
    \eqref{eq:MMSE-usc:MMSE-opt} follows from $g_n$ being the MMSE estimator of $\tY_n$ given $\tX_n$, 
    \eqref{eq:MMSE-usc:triangle-ineq} holds by the triangle (Minkowski) inequality, 
    \eqref{eq:MMSE-usc:epsilons} follows from \eqref{eq:MMSE-usc:tY_n-tY}--\noeqref{eq:Cc_dense_L2}\eqref{eq:hg(Xn)-->hg(X)_m.s.},
    and
    \eqref{eq:MMSE-usc:Skorokhod2} follows from \eqref{eq:MMSE-usc:Skorokhod:CDF:X}.

    Since $k \in \nats$ is fixed and $\eps$ can be chosen to be arbitrarily small, the desired result follows:
    \begin{align}
        \limsup_{n \to \infty} \MMSE{\Y_n}{\X_n} \leq \MMSE{\Y}{\X}. \tag*{\IEEEQED}
    \end{align}


\section{Proof of \lemref{lem:E[Y|Xn]-->E[Y|X]}}
\label{app:proof:lem:E[Y|Xn]-->E[Y|X]}

Denote the dimensions of $\Y$ by $k$, and that of $\X$ (and hence also of $\X_n$) by $m$. 
Define the functions 
\begin{subequations}
\label{eq:MMSE_estimators:g-g_n}
\begin{align}
    g_n\lrp{\x} &\triangleq \CE{\Y}{\X_n = \x},
\label{eq:MMSE_estimator_gn}
 \\ g\lrp{\x} &\triangleq \CE{\Y}{\X = \x}.
\label{eq:MMSE_estimator_g}
\end{align}
\end{subequations}

Since $\Y \markov \X \markov \X_n$ for all $n \in \nats$,
\begin{align}
\label{eq:E[Y|X,Xn]=E[Y|X]}
    \CE{\Y}{\X, \X_n} &= \CE{\Y}{\X} 
\end{align}
a.s.\ for all $n \in \nats$. 
Furthermore, 
\begin{subequations}
\label{eq:proof:g_n(X_n)_eq_E[g(X)|X_n]}
\begin{align}
    g_n(\X_n) &\triangleq \CE{\Y}{\X_n} 
\label{eq:proof:g_n(X_n)_eq_E[g(X)|X_n]:gn}
 \\ &= \CE{\CE{\Y}{\X,\X_n}}{\X_n} 
\label{eq:proof:g_n(X_n)_eq_E[g(X)|X_n]:smooth}
 \\ &= \CE{\CE{\Y}{\X}}{\X_n}    
\label{eq:proof:g_n(X_n)_eq_E[g(X)|X_n]:Markov}
 \\ &= \CE{g\lrp{\X}}{\X_n},     
\label{eq:proof:g_n(X_n)_eq_E[g(X)|X_n]:g}
\end{align}
\end{subequations}
where
\eqref{eq:proof:g_n(X_n)_eq_E[g(X)|X_n]:gn} follows from \eqref{eq:MMSE_estimator_gn},
\eqref{eq:proof:g_n(X_n)_eq_E[g(X)|X_n]:smooth} follows from the law of total expectation, \eqref{eq:proof:g_n(X_n)_eq_E[g(X)|X_n]:Markov} follows from \eqref{eq:E[Y|X,Xn]=E[Y|X]},
and
\eqref{eq:proof:g_n(X_n)_eq_E[g(X)|X_n]:g} follows from~\eqref{eq:MMSE_estimator_g}.\footnote{For $r \geq 2$, 
Equation~\eqref{eq:proof:g_n(X_n)_eq_E[g(X)|X_n]} means that $g_n$ is the MMSE estimator of $g\lrp{\X}$ given $\X_n$.} 

Set $\eps > 0$, however small.
By the same arguments in \appref{app:thm:MMSE-usc} leading to \eqref{eq:Cc_dense_L2} and \eqref{eq:hg(Xn)-->hg(X)_m.s.}, with $L^2$ replaced by $L^r$ and $\lrp{\tX, \tX_n}$ with $\lrp{\X, \X_n}$ (Skorokhod's theorem is not needed since $\X_n \stochto{p} \X$ by assumption) throughout, and $\eps/3$ with $\eps/4$ in \eqref{eq:Cc_dense_L2:i},
there exist $\hg \in C_c \lrp{\reals^m}$ and $n_0$ such that, for all $n > n_0$:
\begin{align} 
\label{eq:MMSE-estimator-continuity:Cc_dense_L2} 
\begin{aligned}
    \normEr{g\lrp{\X} - \hg\lrp{\X}} < \frac{\eps}{4} ,
 \\ \normEr{\hg \lrp{\X} - \hg \lrp{\X_n}} < \frac{\eps}{4}.
\end{aligned}
\end{align}     
In particular, $\normEr{g(\X)}, \normEr{\hg\lrp{\X}}, \normEr{\hg\lrp{\X_n}} < \infty$.

Hence, 
for all $n > n_0$, 
\begin{subequations}
\label{eq:proof:g(X)->g_n(X_n)_m.s.}
\begin{align}
    &\normEr{g\lrp{\X} - g_n(\X_n)}
 \\ &\quad \leq \normEr{g\lrp{\X} - \hg\lrp{\X}} + \normEr{\hg\lrp{\X} - \hg\lrp{\X_n}}
 \\ &\qquad \qquad 
      + \normEr{\hg\lrp{\X_n} - g_n\lrp{\X_n}}
\label{eq:proof:g(X)->g_n(X_n)_m.s.:triangle-ineq1}
 \\ &\quad= \begin{aligned}[t]
      &\normEr{g\lrp{\X} - \hg\lrp{\X}} + \normEr{\hg\lrp{\X} - \hg\lrp{\X_n}}\\
      &\qquad + \normEr{\CE{\hg\lrp{\X_n} - g(\X)}{\X_n}}
    \end{aligned} 
\label{eq:proof:g(X)->g_n(X_n)_m.s.:MMSE-estimator}
 \\ &\quad < \frac{\eps}{2} + \normEr{\hg\lrp{\X_n} - g(\X)} 
\label{eq:proof:g(X)->g_n(X_n)_m.s.:Jensen}
 \\ &\quad\leq \frac{\eps}{2} + \normEr{\hg\lrp{\X_n} - \hg\lrp{\X}}
    + \normEr{\hg\lrp{\X} - g(\X)}
\label{eq:proof:g(X)->g_n(X_n)_m.s.:triangle-ineq2}
 \\ &\quad \leq \eps ,
\label{eq:proof:g(X)->g_n(X_n)_m.s.:eps}
\end{align}
\end{subequations}
where 
\eqref{eq:proof:g(X)->g_n(X_n)_m.s.:triangle-ineq1} and \eqref{eq:proof:g(X)->g_n(X_n)_m.s.:triangle-ineq2} follow from the triangle (Minkowski) inequality;
\eqref{eq:proof:g(X)->g_n(X_n)_m.s.:MMSE-estimator} follows from~\eqref{eq:proof:g_n(X_n)_eq_E[g(X)|X_n]};
\eqref{eq:proof:g(X)->g_n(X_n)_m.s.:Jensen} follows from the (conditional) Jensen inequality~\cite[Chapter~9.7]{Williams:Martingales:Book1991} and the convexity of the norm functional, and from~\eqref{eq:MMSE-estimator-continuity:Cc_dense_L2};
and~\eqref{eq:proof:g(X)->g_n(X_n)_m.s.:eps} follows from~\eqref{eq:MMSE-estimator-continuity:Cc_dense_L2}. Thus, we proved the continuity of the conditional expectation in $L^r$ for $r \geq 1$.
%
%
\hfill \IEEEQED


\section{Proofs of Corollaries \ref{cor:additive-noises} and \ref{cor:machine-precision}}
\label{app:noise-extremes}

\begin{IEEEproof}[Proof of \corref{cor:additive-noises}]
    Let $\seq{\gamma_n}$ and $\seq{\lambda_n \in \reals}$ be some sequences that converge to zero. Denote $\Y_n = \Y + \gamma_n N$ and $\X_n = \X + \lambda_n M$ for $n \in \nats$.

    Since $\normE{\Y}, \normE{N} < \infty$,  
    \begin{align}
        \normE{\Y_n} &\leq \normE{\Y} + \abs{\gamma_n} \normE{N} < \infty &\forall n \in \nats,
    \end{align}
    where the inequality follows from the triangle (Minkowski) inequality.
    Further, 
    \begin{align}
        \limn \normE{\Y_n - \Y} 
        &= \limn \normE{\gamma_n N}
        = \limn \abs{\gamma_n } \cdot \normE{N} = 0
        , 
    \nonumber
    \end{align}
    which means that $\Y_n \stochto{m.s.} \Y$ by \defref{def:convergences}.
    This, in turn implies $\Limn \E{\Y_n^2} = \E{\Y^2}$ as $\normE{\Y} < \infty$.

    Set some $\eps > 0$. Then,
    \begin{align}
        \Limn \PR{\norm{\X_n - \X} > \eps} 
        &= \Limn \PR{\abs{\lambda_n} \cdot \norm{M} > \eps}
        = 0 ,
    \end{align}
    meaning that $\X_n \stochto{p} \X$ by \defref{def:convergences}.
    Hence, 
    \begin{align} 
        \lrp{X_n, Y_n} \stochto{p} \lrp{X,Y}
    \end{align}
    by Lemmata~\ref{lem:convergence:subseq<=>seq} and \ref{lem:convergence:a.s,m.s=>p=>d}.

    Since $M$ is independent of $\lrp{X,Y}$, the Markov condition~\eqref{eq:Markov:intro} 
    \mbox{$\Y \markov \X \markov \X_n$} holds for all $n \in \nats$.
    
    Thus, by \thmref{thm:MMSE:Markov-continuity}, 
    \begin{align}
        \limn \MMSE{\Y + \gamma_n N}{\X + \lambda_n M}
        &= \limn \MMSE{\Y_n}{\X_n} 
     \\ &= \MMSE{\Y}{\X}.
    \end{align}
    Since the above holds for any sequence $\seq{\lrp{\lambda_n, \gamma_n} \in \reals^2}$ that satisfies $\lrp{\lambda_n, \gamma_n} \xto[]{n \to \infty} (0,0)$, the desired result~\eqref{eq:continuity:weak_noise} follows.
\end{IEEEproof}

\begin{IEEEproof}[Proof of \corref{cor:machine-precision}]
    Let $\seq{\gamma_n \in \reals_{>0}}$ and $\seq{\lambda_n \in \reals_{>0}}$ be some sequences of positive reals that converge to zero. Denote $\Y_n = \floor{\Y}_{\gamma_n}$ and $\X_n = \floor{\X}_{\lambda_n}$. Denote by $k$ and $m$ the dimensions of $\Y$ and $\X$, respectively.

    Since $\Y_n - \Y \in \lrp{-\gamma_n, \gamma_n}^k$, $\normE{\Y_n - \Y} < \sqrt{k} \abs{\gamma_n}$. Since $\Limn \gamma_n = 0$, this implies 
    \begin{align}
        \limn \normE{\Y_n - \Y} = 0
        .
    \end{align}
    Since $\normE{\Y} < \infty$, by the triangle (Minkowski) inequality, 
    \begin{align}
        \big| \normE{\Y_n} - \normE{\Y} \big| \leq \normE{\Y_n - \Y}.
    \end{align}
    Thus, $\normE{\Y_n} < \infty$ for all $n \in \nats$ as well.
    Hence, $\Y_n \stochto{m.s.} \Y$ by \defref{def:convergences}.

    Similarly, since $\X_n - \X \in \lrp{-\lambda_n, \lambda_n}^m$, $\norm{\X_n - \X} < \sqrt{m} \abs{\lambda_n}$. Consequently,  
    \begin{align}
        \PR{\norm{\X_n - \X} > \eps} = 0 ,
    \end{align}
    for $\sqrt{m} \abs{\lambda_n} < \eps$. Since $\Limn \lambda_n = 0$, 
    $\X_n \stochto{p} \X$ by \defref{def:convergences}.
    Hence, $\lrp{X_n, Y_n} \stochto{p} \lrp{X,Y}$ by Lemmata~\ref{lem:convergence:subseq<=>seq} and~\ref{lem:convergence:a.s,m.s=>p=>d}.
    
    Since $\X_n = \floor{\X}_{\lambda_n}$ is a deterministic function of $\X$, the Markov condition~\eqref{eq:Markov:intro} 
    \mbox{$\Y \markov \X \markov \X_n$} holds for all $n \in \nats$.
    
    Thus, by \thmref{thm:MMSE:Markov-continuity}, 
    \begin{align}
        \limn \MMSE{\floor{\Y}_{\gamma_n}}{\floor{\X}_{\lambda_n}}
        &= \limn \MMSE{\Y_n}{\X_n} 
     \\ &= \MMSE{\Y}{\X}.
    \end{align}
    Since the above holds for any sequence $\seq{\lrp{\lambda_n, \gamma_n} \in \reals_{>0}^2}$ that satisfies $\lrp{\lambda_n, \gamma_n} \xto[]{n \to \infty} (0,0)$, the desired result~\eqref{eq:continuity:quantization} follows.
\end{IEEEproof}


\section{Proof of \thmref{thm:LMMSE:continuity}}
\label{app:LMMSE}

We will first prove the following lemma.
\begin{lem}
\label{lem:cov-convergence}
    Let $\lrp{X,Y}$ be a pair of RVs and let $\seq{\lrp{X_n, Y_n}}$ be a sequence of pairs of RVs such that
    \begin{itemize}\addtolength\itemsep{.5\baselineskip}
    \item 
        $\lrp{X_n, Y_n} \stochto{d} \lrp{X,Y}$;
    \item 
        $\Limn \E{X_n^2} = \E{X^2} < \infty$;
    \item 
        $\Limn \E{Y_n^2} = \E{Y^2} < \infty$.
    \end{itemize}
    Then, 
    \begin{align}
        \limn \Cov{X_n}{Y_n} = \Cov{X}{Y} ,
    \end{align}
    where $\Cov{A}{B} \triangleq \E{\lrp{A - \E{A}}\lrp{B - \E{B}}}$ denotes the covariance between $A$ and $B$.
\end{lem}

\begin{IEEEproof}
    By Skorokhod's theorem (\thmref{thm:Skorokhod}), there exist $\tilde{X}$, $\tilde{Y}$, $\seq{\lrp{\tilde{X}_n, \tilde{Y}_n}}$ such that
    \begin{subequations}
    \noeqref{eq:LMMSE:Skorokhod:a.s.}
    \label{eq:LMMSE:Skorokhod}
    \begin{align} 
        \lrp{\tilde{X}_n, \tilde{Y}_n}  &\stackrel{d}= \lrp{X_n, Y_n} & \forall n &\in \nats, 
    \label{eq:LMMSE:Skorokhod:CDF:XnYn}
     \\ \lrp{\tilde{X}, \tilde{Y}} &\stackrel{d}= \lrp{X, Y}, 
    \label{eq:LMMSE:Skorokhod:CDF:XY}
     \\ \lrp{\tilde{X}_n, \tilde{Y}_n}  &\stochto{a.s.} \lrp{\tilde{X}, \tilde{Y}} .
    \label{eq:LMMSE:Skorokhod:a.s.}
    \end{align}
    \end{subequations}
    Therefore, 
    \begin{subequations}
    \label{eq:LMMSE:continuity:cov-diff}
    \begin{align}
        &\abs{ \cov{X_n,Y_n} - \cov{X,Y} }
     \\ &\quad= \abs{ \cov{\tilde{X}_n,\tilde{Y}_n} - \cov{\tilde{X},\tilde{Y}} }
    \label{eq:LMMSE:continuity:cov-diff:Skorokhod}
     \\ &\quad = \abs{ \cov{\tilde{X},\tilde{Y}_n-\tilde{Y}} - \cov{\tilde{X} - \tilde{X_n},\tilde{Y_n}} }
    \label{eq:LMMSE:continuity:cov-diff:cov-lin}
     \\ &\quad\leq \begin{aligned}[t]
       &\sqrt{\var{\tilde{X}}}\sqrt{\var{\tilde{Y}_n-\tilde{Y}}}\\
       &\qquad + \sqrt{\var{\tilde{X}_n-\tilde{X}}}\sqrt{{\var{\tilde{Y}_n}}} ,
     \end{aligned}
    \label{eq:LMMSE:continuity:cov-diff:Cauchy-Schwarz}
    \end{align}
    \end{subequations}
    where 
    \eqref{eq:LMMSE:continuity:cov-diff:Skorokhod} follows from \eqref{eq:LMMSE:Skorokhod:CDF:XnYn} and \eqref{eq:LMMSE:Skorokhod:CDF:XY},
    \eqref{eq:LMMSE:continuity:cov-diff:cov-lin} follows from the bilinearity of the covariance,
    and 
    \eqref{eq:LMMSE:continuity:cov-diff:Cauchy-Schwarz} follows from the Cauchy--Schwarz inequality.

    By \thmref{thm:Vitali:prob+2ndMom-->m.s.}, 
    \begin{align}
    \label{eq:LMMSE:continuity:var-diff}
        0 \leq \limn \Var{\tilde{X_n} - \tilde{X}}
        \leq \limn \E{\lrp{\tilde{X_n} - \tilde{X}}^2} 
        = 0,
    \end{align}
    (and similarly for $\tilde{Y_n}$).
    Thus, by the squeeze theorem, 
    \begin{align}
    \label{eq:LMMSE:continuity:var-diff:lim}
    \begin{aligned}
        \limn \Var{\tilde{X_n} - \tilde{X}} &= 0,
     \\ \limn \Var{\tilde{Y_n} - \tilde{Y}} &= 0.
    \end{aligned}
    \end{align}
    Further, by \thmref{thm:Vitali:prob+2ndMom-->m.s.}, 
    \begin{align}
        \limn \Var{\tilde{Y}_n}
        = \Var{\tilde{Y}} < \infty.
    \end{align}

    Hence, the desired result follows from  \eqref{eq:LMMSE:continuity:cov-diff}--\noeqref{eq:LMMSE:continuity:var-diff}\eqref{eq:LMMSE:continuity:var-diff:lim} and the squeeze theorem.
\end{IEEEproof}

We are now ready to prove \thmref{thm:LMMSE:continuity}.

\begin{IEEEproof}[Proof of \thmref{thm:LMMSE:continuity}]
    Since $C_\X$ is invertible, its determinant $\det\lrc{C_\X} \neq 0$ and \thmref{thm:LMMSE} is applicable.

    By \thmref{thm:conv-dist:2ndMom<==>u.i.}, 
    \begin{align}
    \label{eq:LMMSE:means:convergence}
    \begin{aligned}
        \limn \E{X_n} &= \E{X} ,
     \\ \limn \E{Y_n} &= \E{Y} .
     \end{aligned}
    \end{align}

    By \lemref{lem:cov-convergence}, 
    \begin{align}
    \label{eq:cov-matrix:convergence}
    \begin{aligned}
        \limn C_{\X_n} &= C_{\X},
     \\ \limn C_{\Y_n, \X_n} &= C_{\Y, \X}.
    \end{aligned}
    \end{align}
    Hence, by continuity, $\Lim_{n \to \infty} \det\lrc{C_{\X_n}} = \det \lrc{C_{\X}} \neq 0$ meaning that $C_{\X_n}$ is invertible for sufficiently large $n$.
    Furthermore, since 
    \begin{align}
        C_\X^{-1} = \frac{\adj \lrc{C_\X}}{\det \lrc{C_\X}} ,
    \end{align}
    we also have 
    \begin{align}
    \label{eq:inv-cov:convergence}
        \limn C_{\X_n}^{-1} &= C_{\X}^{-1},
    \end{align}
    where $\adj \lrc{C_\X}$ denotes the adjugate of $C_\X$.

    The desired result then follows from the LMMSE formula in \thmref{thm:LMMSE}, \thmref{thm:conv-dist:2ndMom<==>u.i.} 
    and the convergence results in~\eqref{eq:LMMSE:means:convergence}--\noeqref{eq:cov-matrix:convergence}\eqref{eq:inv-cov:convergence}.
\end{IEEEproof}


\section{Proofs for \secref{s:general-cost}}
\label{app:gen-cost:usc:metric-cost}

In this appendix, we denote the dimension of $\Y$ (and hence also of $\Y_n$) by $k$, and that of $\X$ (and hence also of $\X_n$) by $m$. 

\begin{IEEEproof}[Proof of \lemref{lem:Bregman:E[phi(hX_n)]-->E[phi(hX)]}]
    Denote $\hY \triangleq \CE{\Y}{\X}$ and $\hY_n \triangleq \CE{\Y_n}{\X_n}$ for all $n \in \nats$. Set some $\y \in \cY$.

    Since $\lrp{X_n,Y_n} \stochto{p} (X,Y)$, 
    also $\phi\lrp{\Y_n} \stochto{p} \phi(\Y)$ by the continuous mapping theorem (\thmref{thm:continuous-mapping-theorem}). 
    Since also $\Limn \E{\abs{\phi\lrp{\Y_n}}} = \E{\abs{\phi \lrp{\Y}}} < \infty$, 
    $\seq{\phi\lrp{\Y_n}}$ is u.i.\ by Vitali's convergence theorem (\thmref{thm:Vitali:prob+2ndMom-->m.s.}). Then, by \lemref{lem:u.i.-of-conditional-expectation}, $\seq{\CE{\phi\lrp{\Y_n}}{\X_n}}$ is also u.i.
    Taking $\phi$ to be the identity function and repeating the same arguments, with requirement~\ref{itm:Bregman:Markov-continuity:phi-moment-convergence} replaced with requirement~\ref{itm:Bregman:Markov-continuity:moment-convergence}, 
    proves that $\seq{\hY_n}$ is u.i.\ as well. By \thmref{thm:u.i.:properties}, $\seq{\phi\lrp{\y} + \inner{\hY_n - \y}{\nabla \phi \lrp{\y}}}$ is u.i., and hence also 
    $\seq{\abs{\phi\lrp{\y} + \inner{\hY_n - \y}{\nabla \phi \lrp{\y}}} + \abs{\CE{\phi\lrp{\Y_n}}{\X_n}}}$.
    
    Recalling that $\phi$ is convex,
    \begin{align}
        \phi\lrp{\hY_n} &\leq \CE{\phi\lrp{\Y_n}}{\X_n} \qquad a.s. \qquad \forall n \in \nats,
    \end{align}
    by the (conditional) Jensen inequality on the one hand~\cite[Chapter~9.7]{Williams:Martingales:Book1991},
    and 
    \begin{align}
        \phi\lrp{\hY_n} \geq \phi\lrp{\y} + \inner{\hY_n - \y}{\nabla \phi \lrp{\y}}
    \end{align}
    by the subgradient inequality~\cite[Section~23]{Rockafellar:convex-analysis:book1970} on the other hand.
    Consequently, by Property~6 of \thmref{thm:u.i.:properties}, $\seq{\phi\lrp{\hY_n}}$ is u.i.

    By \thmref{thm:Markov-continuity:estimator}, $\hY_n \stochto{L^1} \hY$. Hence, $\hY_n \stochto{p} \hY$ by \lemref{lem:convergence:a.s,m.s=>p=>d}.
    Since $\phi$ is continuous, $\phi \lrp{\hY_n} \stochto{p} \phi \lrp{\hY}$ by the continuous mapping theorem (\thmref{thm:continuous-mapping-theorem}).
    Together with the fact that $\seq{\phi \lrp{\hY_n}}$ is u.i., 
    \begin{align}
        \limn \E{\phi \lrp{\hY_n}} = \E{\phi \lrp{\hY}}
    \end{align}
    by Vitali's convergence theorem (\thmref{thm:Vitali:prob+2ndMom-->m.s.}).
\end{IEEEproof}

To prove \thmref{thm:gen-cost:usc:metric-cost}, we first prove several useful lemmata.

\begin{lem}
\label{lem:app:gen-cost:usc:metric-cost:c(Xn,X)-->0}
    Let $\Y \in \cY$ be an RVec, let $\seq{\Y_n}$ be a sequence of RVecs over a closed subset $\cY \subseteq \reals^k$. 
    Let $\oy \in \cY$, and let $c: \cY \times \cY \to \reals$ be a continuous metric (recall \defref{def:metric-cost}) such that:
    \begin{enumerate}\addtolength\itemsep{.5\baselineskip}
    \item 
    \label{eq:app:gen-cost:usc:metric-cost:c(Xn,X)-->0:Xn-->X}
        $\Y_n \stochto{p} \Y$;
    \item 
    \label{eq:app:gen-cost:usc:metric-cost:c(Xn,X)-->0:E[c(Xn,x)<infty]}
        $\E{\cost{\Y_n}{\oy}}, \E{\cost{\Y}{\oy}} < \infty$ for all $n \in \nats$;
    \item 
    \label{eq:app:gen-cost:usc:metric-cost:c(Xn,X)-->0:E[c(Xn,x)--->E[c(X,x)]}
        $\Limn \E{\cost{\Y_n}{\oy}} = \E{\cost{\Y}{\oy}}$.
    \end{enumerate}
    Then, 
    $\Limn \E{\cost{\Y_n}{\Y}} = 0$.
\end{lem}

\begin{IEEEproof}
    Since $c$ is a continuous mapping and since $\Y_n \stochto{p} \Y$, by the continuous mapping theorem~(\thmref{thm:continuous-mapping-theorem}), we obtain $\cost{\Y_n}{\oy} \stochto{p} \cost{\Y}{\oy}$.
    Under this convergence in probability, Assumptions~\ref{eq:app:gen-cost:usc:metric-cost:c(Xn,X)-->0:E[c(Xn,x)<infty]} and \ref{eq:app:gen-cost:usc:metric-cost:c(Xn,X)-->0:E[c(Xn,x)--->E[c(X,x)]} are equivalent to $\seq{\cost{\Y_n}{\oy}}$ being u.i.\ by Vitali's convergence theorem (\thmref{thm:Vitali:prob+2ndMom-->m.s.}).
    
    By the positivity and triangle inequality properties of a metric: 
    \begin{align}
        0 \leq \cost{\Y_n}{\Y} \leq \cost{\Y_n}{\oy} + \cost{\Y}{\oy};
    \end{align}
    since $\seq{\cost{\Y_n}{\oy}}$ is u.i.\ so is $\cost{\Y_n}{\Y}$.
    Therefore, the desired result follows from Vitali's convergence theorem (\thmref{thm:Vitali:prob+2ndMom-->m.s.}).
\end{IEEEproof}

To prove the next lemma we first recall the following two theorems adapted for probability over finite-dimensional Euclidean spaces.
\begin{thm}[Lusin's theorem~{\cite[Theorem~1.14]{Evans-Gariepy:Measure-theory:Book2025}}]
\label{thm:Lusin}
    Let $\X \in \cX \subseteq \reals^m$ be a RVec.
    Let $f: \cX \to \reals^k$ be a measurable function.
    Let $\eps > 0$, however small. 
    Then, there exists a compact set $\cS \subseteq \cX$, such that 
    \begin{align}
    \label{eq:Lusin:compact-support}
        \PR{\X \notin \cS} &< \eps
    \end{align}
    and the restriction $f|_S$ of $f$ to $\cS$ is continuous.
\end{thm}
\begin{thm}[Tietze--Urysohn--Brouwer--Dugundji extension~{\cite{Dugundji1951:generalized-Tietze-extension:1951}}]
\label{thm:Tietze}
    Let $\cS$ be a non-empty closed subset of $\reals^m$ and let $f: \cS \to \reals^k$ be continuous.
    Then, there exists a continuous map $g: \reals^m \to \reals^k$ such that $g(\y) = f(\y)$ for all $\y \in \cS$ and $g\lrp{\reals^m} \subseteq \textrm{convex hull of } f(\cS)$. 
\end{thm}


\begin{lem}
\label{lem:app:gen-cost:usc:metric-cost:approx-g}
    Let $\X$ be an RVec over $\cX \subseteq \reals^m$, and let $\cY \subset \reals^k$ be a closed convex set.
    Let $c: \cY \times \cY \to \reals_{\geq 0}$ be a continuous function satisfying $c(x,x) = 0$ for all $\y \in \cY$.
    Let $f: \reals_{\geq 0} \to \reals_{\geq 0}$ be a continuous strictly increasing function
    that satisfies $f(0) = 0$ and assume that $c$ satisfies a generalized triangle inequality with respect to~f as in~\corref{cor:gen-cost:usc:metric-cost-after-func}.
    Let $g: \cX \to \cY$ be a measurable function such that $\E{\cost{g\lrp{\X}}{\oy}} < \infty$ for some constant $\oy \in \cY$.
    Then:
    \begin{enumerate}
    \item 
        There exist constants $\alpha, \beta >0$ that depend only on $f$, such that 
        \begin{align}
            \cost{\y_1}{\y_3} \leq \alpha \lrs{\cost{\y_1}{\y_2} + \cost{\y_2}{\y_3}} + \beta
        \end{align}
        for all $\y_1, \y_2, \y_3 \in \cY$.
    \item 
        For every $\eps > 0$, 
        there exists a continuous bounded function $\hg: \cX \to \cY$, 
        such that 
        \begin{align}
        \label{eq:app:gen-cost:usc:metric-cost:approx-g}
            \E{\cost{g\lrp{\X}}{\hg\lrp{\X}}} &< \eps .
        \end{align}
    \end{enumerate}
\end{lem}

\begin{IEEEproof}
    We first establish an affine bound on $c$.
    Let $\beta > 0$ and define 
    \begin{align}
    \label{eq:metrix:affine-approx:b-def}
        b &= f^{-1} \lrp{\frac{f(\beta)}{2}}, & \alpha &= \frac{\beta}{b}.
    \end{align}
    Since $f$ is continuous and strictly monotone it is also invertible. Further, since $f(0) = 0$, the value $f(\beta)/2$ has a unique preimage $b \in (0, \beta)$. Thus, $b$ is well-defined and $2f(b) = f(\beta)$.

    Now, fix some $\y_1, \y_2, \y_3 \in \cY$, and set
    \begin{align}
    \label{eq:metrix:affine-approx:p-def}
        p = \frac{b}{\max \lrc{b, \cost{\y_1}{\y_2}, \cost{\y_2}{\y_3}}}.
    \end{align}
    Clearly $0 < p \leq 1$, and 
    \begin{align}
    \label{eq:pc<b}
        p \cdot \cost{\y_1}{\y_2} &\leq b, & p \cdot \cost{\y_2}{\y_3} \leq b.
    \end{align}
    Let $B$ be a Bernoulli RV with parameter $p$, and define 
    \begin{align}
        A_i &= B \y_i + (1-B) \oy, & i \in \{1,2,3\}.
    \end{align}
    Since $\cost{\oy}{\oy} = 0$, 
    \begin{align}
    \label{eq:metric:affine-approx:Bernoulli-def}
        \E{\cost{A_i}{A_j}} &= p \cost{\y_i}{\y_j} < \infty &\forall i,j &\in \{1,2,3\}.
    \end{align}
    Therefore, 
    \begin{subequations}
    \label{eq:metric:affine-approx}
    \begin{align}
        f\lrp{p \cost{\y_1}{\y_3}} &= f \lrp{\E{\cost{A_1}{A_3}}} 
    \label{eq:metric:affine-approx:Bernoulli1}
     \\ & \leq f \lrp{\E{\cost{A_1}{A_2}}} + f \lrp{\E{\cost{A_2}{A_3}}} \qquad
    \label{eq:metric:affine-approx:triangle-ineq}
     \\ & = f\lrp{p \cost{\y_1}{\y_2}} + f\lrp{p \cost{\y_2}{\y_3}} 
    \label{eq:metric:affine-approx:Bernoulli2}
     \\ & \leq 2 f(b) 
    \label{eq:metric:affine-approx:p-def-use}
     \\ & = f(\beta) ,
    \label{eq:metric:affine-approx:b-beta}
    \end{align}
    \end{subequations}
    where 
    \eqref{eq:metric:affine-approx:Bernoulli1} and~\eqref{eq:metric:affine-approx:Bernoulli2} follow from~\eqref{eq:metric:affine-approx:Bernoulli-def},
    \eqref{eq:metric:affine-approx:triangle-ineq} follows from the generalized triangle inequality,
    \eqref{eq:metric:affine-approx:p-def-use} follows from~\eqref{eq:pc<b} since $f$ is increasing, 
    and 
    \eqref{eq:metric:affine-approx:b-beta} follows from~\eqref{eq:metrix:affine-approx:b-def}.
    
    Furthermore,  
    \begin{subequations}
    \label{eq:cost:affine-approx:final}
    \begin{align}
        \cost{\y_1}{\y_3} &\leq \frac{\beta}{p} 
    \label{eq:cost:affine-approx:final:monotone}
     \\ &= \alpha \max \lrc{b, \cost{\y_1}{\y_2}, \cost{\y_2}{\y_3}}
    \label{eq:cost:affine-approx:final:alpha-def1}
     \\ &\leq \alpha \lrs{b + \cost{\y_1}{\y_2} + \cost{\y_2}{\y_3}}
    \label{eq:cost:affine-approx:final:max<sum}
     \\ &= \alpha \lrs{\cost{\y_1}{\y_2} + \cost{\y_2}{\y_3}} + \beta ,
    \label{eq:cost:affine-approx:final:alpha-def2}
    \end{align}
    \end{subequations}
    where
    \eqref{eq:cost:affine-approx:final:monotone} holds since $f$ is increasing and invertible,
    \eqref{eq:cost:affine-approx:final:alpha-def1} and~\eqref{eq:cost:affine-approx:final:alpha-def2} follow from~\eqref{eq:metrix:affine-approx:b-def} and~\eqref{eq:metrix:affine-approx:p-def}, 
    and
    \eqref{eq:cost:affine-approx:final:max<sum} holds since all the elements in the maximum are non-negative.
    
    Let $\eps > 0$ and choose $\eta > 0$ such that 
    \begin{align}
    \label{eq:eta-def}
        2 f(\eta) < f(\eps).
    \end{align}
    For $\tau \in \reals$ satisfying $\tau > \normInf{\oy}$, define 
        \begin{align}
        \label{eq:g_r}
        g_\tau\lrp{\x} = 
        \begin{cases}
            g\lrp{\x}, & \normInf{g\lrp{\x}} \leq \tau
         \\ \oy, & \mathrm{otherwise}
        \end{cases}
    \end{align}
    for $\x \in \cX$. 
    In particular, $g_\tau(\x) \in \cY_\tau \triangleq \cY \cap \lrcm{\y\, }{\, \norm{\y}_\infty \leq \tau}$ for all $\x \in \cX$ and $\cY_\tau$ is compact and convex as the intersection of a closed convex set ($\cY$) and a compact convex set ($\ell_\infty$-ball of radius $\tau$).
    
    Since the cost $c$ is continuous on $\cY \times \cY$, 
    \begin{align}
    \label{eq:proof:generic-cost:usc:max_cost}
        c^{\max}_\tau \triangleq \max_{\y_1, \y_2 \in \cY_\tau} \cost{\y_1}{\y_2} < \infty.
    \end{align}
    
    Since $\E{\cost{g\lrp{\X}}{\oy}} < \infty$, 
    there exists $\tau > 0$, large enough, such that 
    \begin{align}
    \label{eq:proof:generic-cost:usc:cost-of-truncation}
        \E{\cost{g\lrp{\X}}{g_\tau\lrp{\X}}} &= \int\limits_{\x:\, \normInf{g \lrp{\x}} > \tau} 
        \!\!\!\!\!\!\!\! 
        \cost{g\lrp{\x}}{\oy} P(dy) 
        < \eta \, . \ \ 
    \end{align}

    By \thmref{thm:Lusin}, there exists a compact set $\cS \subseteq \cX$, such that 
    \begin{align}
    \label{eq:proof:app:gen-cost:usc:metric-cost:approx-g:Lusin:compact-support}
        \PR{\X \notin \cS} &< \min \lrc{\half, \frac{\eta}{1 + c^{\max}_\tau}}, 
    \end{align}
    and $\left. g_\tau\right|_\cS$ is continuous.

    By \thmref{thm:Tietze} and since $\cS$ is closed as a compact set, there exists a continuous map $\hg: \reals^m \to \reals^k$ such that 
    \begin{align}
    \label{eq:proof:app:gen-cost:usc:metric-cost:approx-g:Tietze}
    \begin{aligned}
        \hg(\x) &= g_\tau(\x) & \forall \x &\in \cS,
     \\ \hg \lrp{\tilde{\x}} &\in \cY_\tau & \forall \tilde{\x} &\in \cX,
    \end{aligned}
    \end{align}
    where we used the fact that $\cY_\tau$ is convex and hence equals its convex hull.

    By \eqref{eq:proof:generic-cost:usc:max_cost} and~\eqref{eq:proof:app:gen-cost:usc:metric-cost:approx-g:Lusin:compact-support}, we have 
    \begin{align}
        \begin{aligned}
        \E{\cost{g_\tau \lrp{\X}}{\hg \lrp{\X}}}
        &= \int\limits_{\x \notin \cS} \cost{g_\tau \lrp{\x}}{\hg \lrp{\x}} \PR{d\x}\\
        &\leq c^{\max}_\tau \cdot \PR{Y \notin \cS} < \eta.
        \end{aligned}
    \label{eq:proof:generic-cost:usc:cost-of-continuity}
    \end{align}

    By applying \eqref{eq:cost:affine-approx:final}, \eqref{eq:proof:generic-cost:usc:cost-of-truncation}, and \eqref{eq:proof:generic-cost:usc:cost-of-continuity} we obtain
    \begin{align}
        &\E{\cost{g \lrp{Y}}{\hg \lrp{Y}}}
     \\ &\quad \leq \alpha \lrp{ \E{\cost{g \lrp{Y}}{g_\tau \lrp{Y}}}
          + \E{\cost{g_\tau \lrp{Y}}{\hg \lrp{Y}}}} + \beta
     \\ &\quad< 2 \alpha \eta + \beta < \infty,
    \end{align}
    meaning that all three expectations are finite. 
    Hence, 
    \begin{subequations}
    \label{eq:proof:generic-cost:final-step}
    \begin{align}
        &f\lrp{ \E{\cost{g \lrp{Y}}{\hg \lrp{Y}}} }
     \\ &\quad\leq 
          f \lrp{ \E{\cost{g \lrp{Y}}{g_\tau \lrp{Y}}} }
          + f\lrp{ \E{\cost{g_\tau \lrp{Y}}{\hg \lrp{Y}}} } \qquad 
    \label{eq:proof:generic-cost:final-step:triangle-ineq}
     \\ &\quad < 2 f(\eta) 
    \label{eq:proof:generic-cost:final-step:eta-bound}
     \\ &\quad < f(\eps) ,
    \label{eq:proof:generic-cost:final-step:eta<eps}
    \end{align}
    \end{subequations}
    where 
    \eqref{eq:proof:generic-cost:final-step:triangle-ineq} follows from the generalized triangle inequality,
    \eqref{eq:proof:generic-cost:final-step:eta-bound} follows from \eqref{eq:proof:generic-cost:usc:cost-of-truncation} and \eqref{eq:proof:generic-cost:usc:cost-of-continuity},
    and~\eqref{eq:proof:generic-cost:final-step:eta<eps} follows from~\eqref{eq:eta-def}.
    Recalling that $f$ is increasing and invertible proves the desired result~\eqref{eq:app:gen-cost:usc:metric-cost:approx-g}.
\end{IEEEproof}

We are now ready to prove \thmref{thm:gen-cost:usc:metric-cost}.
\begin{IEEEproof}[Proof of \thmref{thm:gen-cost:usc:metric-cost}]
    Let $\eps > 0$.
    Let $g^*$ be a measurable function such that 
    \begin{align}
    \label{eq:g^*-choice}
        \E{\cost{\Y}{g^*\lrp{\X}}} \leq \inf_{g} \E{\cost{\Y}{g\lrp{\X}}} + \eps,
    \end{align}
    where the infimum is taken over all measurable functions $g: \cX \to \cY$.
    Note that, by the triangle inequality, 
    \begin{align}
    \label{eq:g*}
        \E{\cost{g^*\lrp{\X}}{\oy}}
        &\leq \E{\cost{\Y}{g^*\lrp{\X}}} + \E{\cost{\Y}{\oy}} < \infty.\quad\ 
    \end{align}
    
    Now, by \lemref{lem:app:gen-cost:usc:metric-cost:approx-g} applied with $f(t) = t$, there exists a continuous bounded function $\hg$ such that 
    \begin{subequations}
    \begin{align}
     \\ \E{\cost{g^*\lrp{\X}}{\hg\lrp{\X}}} &< \eps.
    \label{eq:c(g-approx,g)<eps}
    \end{align}
    \end{subequations}

    By Skorokhod's theorem (\thmref{thm:Skorokhod}), there exist $\tilde{X}$, $\tilde{Y}$, $\seq{\tilde{X}_n}$, and $\seq{\tilde{Y}_n}$ that satisfy:
    \begin{subequations}
    \label{eq:gen-cost-usc:Skorokhod}
    \noeqref{eq:gen-cost-usc:Skorokhod:a.s.}
    \begin{align} 
        \lrp{\tilde{X}_n, \tilde{Y}_n} &\stackrel{d}= \lrp{X_n, Y_n} & \forall n &\in \nats, 
    \label{eq:gen-cost-usc:Skorokhod:CDF:Xn}
     \\ \lrp{\tilde{X}, \tilde{Y}} &\stackrel{d}= \lrp{X, Y}, 
    \label{eq:gen-cost-usc:Skorokhod:CDF:X}
     \\ \lrp{\tilde{X}_n, \tilde{Y}_n} &\stochto{a.s.} \lrp{\tilde{X}, \tilde{Y}}.
    \label{eq:gen-cost-usc:Skorokhod:a.s.}
    \end{align}
    \end{subequations}    
    In particular, $\lrp{\tilde{X}_n, \tilde{Y}_n} \stochto{p} \lrp{\tilde{X}, \tilde{Y}}$ by \lemref{lem:convergence:a.s,m.s=>p=>d}, and 
    \begin{align}
    \label{eq:gen-cost-usc:E[Xn,X)]-->0}
        \lim_{n \to \infty} \E{\cost{\tY_n}{\tY}} = 0, 
    \end{align}
    by \lemref{lem:app:gen-cost:usc:metric-cost:c(Xn,X)-->0}.

    Since $\tX_n \stochto{p} \tX$, and $\hg$ and $c$ are continuous, $\hg \lrp{\tX_n} \stochto{p} \hg \lrp{\tX}$ and $\cost{\hg(\tX_n)}{\oy} \stochto{p} \cost{\hg(\tX)}{\oy}$ by the continuous mapping theorem~(\thmref{thm:continuous-mapping-theorem}). 
    Since $\hg$ is bounded and $\cY$ is closed, the closure of its range is a compact subset of $\cY$. Continuity of $c$ makes these costs uniformly bounded and hence u.i.\ by 
    Theorems~\ref{thm:u.i.:properties} and \ref{thm:Vitali:prob+2ndMom-->m.s.}:
    \begin{align}
    \label{eq:usc:metric-cost:proof:compact-c(g(Yn),x)-->c(g(Y,x))}
        \cost{\hg \lrp{\tX_n}}{\oy} \stochto{L^1} \cost{\hg \lrp{\tX}}{\oy}.
    \end{align}
    Now, by again appealing to \lemref{lem:app:gen-cost:usc:metric-cost:c(Xn,X)-->0}, we obtain
    \begin{align}
    \label{eq:gen-cost-usc:E[hg(Yn),hg(Y))]-->0}
        \lim_{n \to \infty} \E{\cost{\hg \lrp{\tX_n}}{\hg \lrp{\tX}}} = 0.
    \end{align} 

    Therefore, there exists $n_0 \in \nats$ such that, for all $n > n_0$, 
    \begin{subequations}
    \label{eq:usc:metric-cost:proof}
    \begin{align}
         &\ocost{\Y_n}{\X_n}
    \\ &\qquad = \inf_{g:\ \cX \to \cY} \E{c \lrp{\Y_n, g\lrp{\X_n}}} 
    \label{eq:usc:metric-cost:proof:opt-cost-def1}
     \\ &\qquad \leq \E{ \cost{\Y_n}{\hg\lrp{\X_n}} }
    \label{eq:usc:metric-cost:proof:opt-g}
     \\ &\qquad = \E{ \cost{\tY_n}{\hg\lrp{\tX_n}} }
    \label{eq:usc:metric-cost:proof:Skorokhod1}
     \\ &\qquad\leq 
        \E{ \cost{\tY_n}{\tY} } + \E{ \cost{\tY}{g^*\lrp{\tX}} }
     \\ &\qqquad + \E{ \cost{g^*\lrp{\tX}}{\hg\lrp{\tX}} }
       + \E{ \cost{\hg\lrp{\tX}}{\hg\lrp{\tX_n}} }\quad \ \ 
    \label{eq:usc:metric-cost:proof:triangle-ineq}
     \\ &\qquad\leq \E{ \cost{\tY}{g^*\lrp{\tX}} }
       + \E{ \cost{g^*\lrp{\tX}}{\hg\lrp{\tX}} } + 2\eps
    \label{eq:usc:metric-cost:proof:<eps}
     \\ &\qquad= \E{ \cost{\Y}{g^*\lrp{\X}} }
       + \E{ \cost{g^*\lrp{\X}}{\hg\lrp{\X}} } + 2\eps
    \label{eq:usc:metric-cost:proof:Skorokhod2}
     \\ &\qquad \leq \inf_{g:\ \cX \to \cY} \E{ \cost{\Y}{g\lrp{\X}} } + 4\eps 
    \label{eq:usc:metric-cost:proof:g*-approx}
     \\ &\qquad = \ocost{\Y}{\X} + 4\eps ,
    \label{eq:usc:metric-cost:proof:opt-cost-def2}
    \end{align}
    \end{subequations}
    where 
    \eqref{eq:usc:metric-cost:proof:opt-cost-def1} and \eqref{eq:usc:metric-cost:proof:opt-cost-def2} follow from \eqref{eq:general-cost:optimal-cost-def};
    \eqref{eq:usc:metric-cost:proof:opt-g} holds since $\hg$ is a measurable admissible estimator;
    \eqref{eq:usc:metric-cost:proof:Skorokhod1} follows from \eqref{eq:gen-cost-usc:Skorokhod:CDF:Xn};
    \eqref{eq:usc:metric-cost:proof:triangle-ineq} follows from the triangle inequality;
    \eqref{eq:usc:metric-cost:proof:<eps} follows from \eqref{eq:gen-cost-usc:E[Xn,X)]-->0} and \eqref{eq:gen-cost-usc:E[hg(Yn),hg(Y))]-->0} for a sufficiently large $n_0 \in \nats$;
    \eqref{eq:usc:metric-cost:proof:Skorokhod2} follows from \eqref{eq:gen-cost-usc:Skorokhod:CDF:X};
    and \eqref{eq:usc:metric-cost:proof:g*-approx} follows from \eqref{eq:g^*-choice} and \eqref{eq:c(g-approx,g)<eps}.
    Recalling that $\eps > 0$ can be chosen to be arbitrarily small concludes the proof.
\end{IEEEproof}

We now explain how to modify the proof of \thmref{thm:gen-cost:usc:metric-cost} to attain \corref{cor:gen-cost:usc:metric-cost-after-func}. 

\begin{IEEEproof}[Proof of {\corref{cor:gen-cost:usc:metric-cost-after-func}}]
    Let $\delta > 0$. Since $f$ is continuous at zero and $f(0) = 0$, choose $\eps > 0$ such that $f(\eps) < \delta$.
    Since $f$ is increasing, this also guarantees that
    \begin{align}
    \label{eq:gen-cost:usc:generalized-triangle-ineq:eps<-->delta}
        0 \leq \tilde{\eps} < \eps \quad \Leftrightarrow \quad 0 \leq f\lrp{\tilde{\eps}} < f(\eps) < \delta.
    \end{align}

    Let $g^*$ be a measurable function with 
    $\E{\cost{\Y}{g^*(\X)}} < \infty$
    such that 
    \begin{align}
    \label{eq:g^*-choice:f()}
        f\lrp{ \E{\cost{\Y}{g^*\lrp{\X}}} } \leq \inf_{g} f\lrp{ \E{\cost{\Y}{g\lrp{\X}}} } + \delta.
    \end{align}
    Here and in the remainder of the proof, each infimum is taken over all measurable functions $g: \cX \to \cY$ for which the corresponding mean cost is finite.
    
    By \lemref{lem:app:gen-cost:usc:metric-cost:approx-g} and the symmetry of $c$, 
    \begin{align}
        \begin{aligned}
        \E{\cost{g^*\lrp{\X}}{\oy}}
        &\leq \alpha \lrp{ \E{\cost{\Y}{g^*\lrp{\X}}} + \E{\cost{\Y}{\oy}} } + \beta\\
        &< \infty,
        \end{aligned}
    \label{eq:g*:f()}
    \end{align}
    for constants $\alpha, \beta > 0$ that depend only on $f$.
    Further by \lemref{lem:app:gen-cost:usc:metric-cost:approx-g} with the present function $f$, choose a continuous bounded function $\hg$ satisfying~\eqref{eq:c(g-approx,g)<eps}.
    
    By Skorokhod's theorem (\thmref{thm:Skorokhod}), there exist $\tilde{X}$, $\tilde{Y}$, $\seq{\tilde{X}_n}$, and $\seq{\tilde{Y}_n}$ that satisfy \eqref{eq:gen-cost-usc:Skorokhod}.
    Define
    \begin{align}
        Z_n &\triangleq \cost{\tY_n}{\oy}, & Z &= \cost{\tY}{\oy}.
    \end{align}
    By the continuous mapping theorem (\thmref{thm:continuous-mapping-theorem}) 
    $Z_n \stochto{a.s.} Z$, since $c$ is continuous and~\eqref{eq:gen-cost-usc:Skorokhod:a.s.}.
    By~\eqref{eq:gen-cost-usc:Skorokhod:CDF:Xn}, \eqref{eq:gen-cost-usc:Skorokhod:CDF:X}, and the last two assumptions of the corollary, 
    \begin{align}
        \lim_{n \to \infty} \E{Z_n} = \lim_{n \to \infty}\E{\cost{\tY_n}{\oy}} = \E{\cost{\tY}{\oy}} = \E{Z} < \infty.
    \end{align}
    Since $Z_n \stochto{a.s.} Z$ (and hence also $Z_n \stochto{p} Z$), 
    Vitali's convergence theorem (\thmref{thm:Vitali:prob+2ndMom-->m.s.}) with $r=1$ implies that $Z_n \stochto{L^1} Z$ and that $\seq{Z_n}$ is u.i.
    By the affine bound of \lemref{lem:app:gen-cost:usc:metric-cost:approx-g} and symmetry, 
    \begin{align}
        0 \leq \cost{\tY_n}{\tY} \leq \alpha \lrp{Z_n + Z} + \beta.
    \end{align}
    Thus $\seq{\cost{\tY_n}{\tY}}$ is u.i.\ by \thmref{thm:u.i.:properties}.
    Moreover, the continuous mapping theorem (\thmref{thm:continuous-mapping-theorem}), \eqref{eq:gen-cost-usc:Skorokhod:a.s.} and $\cost{\y}{\y} = 0$ give
    \begin{align}
        \cost{\tY_n}{\tY} \stochto{a.s.} 0.
    \end{align}
    Thus, \lemref{lem:convergence:a.s,m.s=>p=>d} and Vitali's convergence theorem (\thmref{thm:Vitali:prob+2ndMom-->m.s.}) establish~\eqref{eq:gen-cost-usc:E[Xn,X)]-->0}.

    Similarly, \eqref{eq:gen-cost-usc:E[hg(Yn),hg(Y))]-->0} also holds here by the continuous mapping theorem, the same compactness argument, and Vitali’s convergence theorem.



    Then, there exists $n_0 \in \nats$ such that, for all $n > n_0$, 
    \begin{subequations}
    \label{eq:usc:f(metric-cost):proof}
    \begin{align}
        &\inf_{g} f\lrp{ \E{ \cost{\Y_n}{g\lrp{\X_n}} } }
        \leq f\lrp{ \E{ \cost{\Y_n}{\hg\lrp{\X_n}} } }
    \label{eq:usc:f(metric-cost):proof:opt-g}
     \\ & = f\lrp{ \E{ \cost{\tY_n}{\hg\lrp{\tX_n}} } }
    \label{eq:usc:f(metric-cost):proof:Skorokhod1}
     \\ &\leq 
       f\lrp{ \E{ \cost{\tY_n}{\tY} } }
       + f\lrp{ \E{ \cost{\tY}{g^*\lrp{\tX}} } }
      \\ &\quad + f\lrp{ \E{ \cost{g^*\lrp{\tX}}{\hg\lrp{\tX}} } }
       + f\lrp{ \E{ \cost{\hg\lrp{\tX}}{\hg\lrp{\tX_n}} } } \qquad
    \label{eq:usc:f(metric-cost):proof:triangle-ineq}
     \\ &\leq f\lrp{ \E{ \cost{\tY}{g^*\lrp{\tX}} } }
       + f\lrp{ \E{ \cost{g^*\lrp{\tX}}{\hg\lrp{\tX}} } } + 2\delta
    \label{eq:usc:f(metric-cost):proof:<eps}
     \\ &= f\lrp{ \E{ \cost{\Y}{g^*\lrp{\X}} } }
       + f\lrp{ \E{ \cost{g^*\lrp{\X}}{\hg\lrp{\X}} } } + 2\delta
    \label{eq:usc:f(metric-cost):proof:Skorokhod2}
     \\ &\quad \leq \inf_{g} f \lrp{\E{ \cost{\tY}{g\lrp{\tX}} }} + 4\delta 
    \label{eq:usc:f(metric-cost):proof:g*-approx}
    \end{align}
    \end{subequations}
    where 
    \eqref{eq:usc:f(metric-cost):proof:opt-g} holds since $\hg$ is a measurable admissible estimator with finite expected cost, as established above;
    \eqref{eq:usc:f(metric-cost):proof:Skorokhod1} follows from \eqref{eq:gen-cost-usc:Skorokhod:CDF:Xn};
    \eqref{eq:usc:f(metric-cost):proof:triangle-ineq} follows from the generalized triangle inequality;
    \eqref{eq:usc:f(metric-cost):proof:<eps} follows from \eqref{eq:gen-cost-usc:E[Xn,X)]-->0}, \eqref{eq:gen-cost-usc:E[hg(Yn),hg(Y))]-->0}, and \eqref{eq:gen-cost:usc:generalized-triangle-ineq:eps<-->delta} for a sufficiently large $n_0 \in \nats$;
    \eqref{eq:usc:f(metric-cost):proof:Skorokhod2} follows from \eqref{eq:gen-cost-usc:Skorokhod:CDF:X};
    and
    \eqref{eq:usc:f(metric-cost):proof:g*-approx} follows from \eqref{eq:c(g-approx,g)<eps}, \eqref{eq:g^*-choice:f()}, and \eqref{eq:gen-cost:usc:generalized-triangle-ineq:eps<-->delta}.
    By the integrability assumptions, \eqref{eq:gen-cost-usc:Skorokhod}, \eqref{eq:g*:f()}, and boundedness of $\hat g$, all the RVecs in~\eqref{eq:usc:f(metric-cost):proof} have finite mean costs to $\oy$. In particular, the affine bound of \lemref{lem:app:gen-cost:usc:metric-cost:approx-g} and symmetry of $c$ ensure that all means required for \eqref{eq:usc:f(metric-cost):proof:triangle-ineq} are finite.

    Recalling that $\delta > 0$ can be chosen to be arbitrarily small and that $f$ is a strictly increasing continuous function that satisfies $f(0) = 0$, and hence also the inverse function thereof, concludes the proof.
\end{IEEEproof}


\section*{Acknowledgments}

The authors thank Pavel Chigansky for helpful discussions about the proof of \thmref{thm:MMSE:Markov-continuity} and uniform integrability, Amir Puri for an interesting discussion about the proper formulation of Markovian continuity, Aya Vituri for providing feedback and comments, and Ido Nachum for valuable discussions about the problem definition in machine learning contexts and about Skorokhod's theorem in its general form.





\end{document}